\documentclass[preprint,showpacs,nofootinbib,eqsecnum,amsfonts,amsmath,aps]{revtex4}
 
\allowdisplaybreaks

\begin{document}

\title{Third post-Newtonian dynamics of compact binaries:\\Equations
of motion in the center-of-mass frame}

\author{Luc Blanchet}\email{blanchet@iap.fr}\altaffiliation{On leave from Institut d'Astrophysique
de Paris, groupe de Gravitation et Cosmologie (GR$\varepsilon$CO, FRE
2435 du CNRS), 98$^{\rm bis}$ boulevard Arago, 75014 Paris, France}
\affiliation{Mc Donnell Center for the Space Sciences,\\ Department of
Physics, Washington University,\\ St. Louis, Missouri 63130, U.S.A.}
\author{Bala R. Iyer}\email{bri@rri.res.in}\affiliation{Raman Research Institute,\\ Bangalore
560080, India}

\date{\today}

\begin{abstract}
The equations of motion of compact binary systems and their associated
Lagrangian formulation have been derived in previous works at the
third post-Newtonian (3PN) approximation of general relativity in
harmonic coordinates. In the present work we investigate the binary's
relative dynamics in the center-of-mass frame (center of mass located
at the origin of the coordinates). We obtain the 3PN-accurate
expressions of the center-of-mass positions and equations of the
relative binary motion. We show that the equations derive from a
Lagrangian (neglecting the radiation reaction), from which we deduce
the conserved center-of-mass energy and angular momentum at the 3PN
order. The harmonic-coordinates center-of-mass Lagrangian is
equivalent, {\it via} a contact transformation of the particles'
variables, to the center-of-mass Hamiltonian in ADM coordinates that
is known from the post-Newtonian ADM-Hamiltonian formalism. As an
application we investigate the dynamical stability of circular binary
orbits at the 3PN order.
\end{abstract}

\maketitle

\section{Introduction}

The problem of the dynamics of two compact bodies is part of a larger
program aimed at unraveling the information contained in the
gravitational-wave signals emitted by inspiralling and/or coalescing
compact binaries (see Refs. \cite{W94,Blivrev} for reviews). The
current treatment of the problem is Post-Newtonian (expansion when the
speed of light $c\to +\infty$; following the standard practice we say
that a term of order $1/c^{2n}$ relative to the Newtonian force
belongs to the $n$PN approximation). The first breakthrough in the
problem of dynamics has been the completion of the equations of motion
of two point-like particles up to the 2PN order
\cite{DD81a,DD81b,D82,DS85,Kop85,BFP98,IFA01,PW02}. In recent years the (quite
involved) equations of motion at the next 3PN order have also been
successfully derived
\cite{JaraS98,JaraS99,DJS00,DJS01,DJSdim,BF00,BFeom,BFreg,BFregM,ABF01}.

Up to the order 3.5PN there is a clean separation between the
conservative part of the dynamics, made of the ``even'' Newtonian,
1PN, 2PN and 3PN approximations --- with the 2PN and especially the
3PN ones being very difficult to obtain ---, and the part associated
with the radiation reaction, and consisting of the ``odd'' 2.5PN and
3.5PN orders (which are comparatively much simpler to control than 2PN
and 3PN). In principle the conservative part of the equations yields a
point of dynamical general-relativistic instability at which there is
(presumably) a transition from the adiabatic inspiral to the final
plunge and coalescence. On the other hand, the non-conservative terms
--- i.e. 2.5PN computed in
Refs. \cite{DD81a,DD81b,D82,DS85,Kop85,BFP98}, 3.5PN and 4.5PN
computed in Refs. \cite{IW93,IW95,GII97,PW02} --- are determined by
the boundary conditions imposed on the gravitational field at
infinity.

One should not confuse the latter nomenclature for post-Newtonian
orders with a different one applied to the gravitational field at
future null infinity. There the ``Newtonian'' order, which has a
quadrupolar wave pattern, corresponds to the dominant odd term in the
local equations of motion, i.e. 2.5PN, while the 1PN order, which is
both quadrupolar and octupolar, corresponds to 3.5PN in the local
equations. And so on. Because of the presence of tails at the 1.5PN
order in the wave field at infinity there is a contribution at the 4PN
order in the equations of motion that is ``odd'' in the sense of being
associated with radiation-reaction effects \cite{BD88}. Similarly one
expects that the known tails-of-tails \cite{B98tail} arising at the
3PN order in the wave field will correspond to an odd contribution at
the 5.5PN order in the equations of motion\footnote{This difference by
2.5PN orders explains why the equations of motion are insufficient as
regards the radiative aspects of the problem. For analyzing the waves
emitted by inspiralling compact binaries one needs not only the
solution of the problem of motion but also the (equally crucial)
solution of the problem of gravitational-wave generation
\cite{Blivrev}.}.

Two different methods, relying on two independent frameworks, have
been applied to the equations of motion at the 3PN order. Jaranowski
and Sch\"afer \cite{JaraS98,JaraS99}, working within the
ADM-Hamiltonian formalism of general relativity, derived the
Hamiltonian describing the motion of two compact bodies, in ADM
coordinates, and in the center-of-mass frame. The Hamiltonian was
later generalized to an arbitrary frame in Ref. \cite{DJS00}. Blanchet
and Faye \cite{BF00,BFeom,BFreg,BFregM} (following the method proposed
in Ref. \cite{BFP98}) performed a direct iteration of the equations of
motion in harmonic coordinates, and in a general frame. The Lagrangian
of the motion was then deduced from the equations of motion
\cite{ABF01}. The end results provided by these two methods ---
ADM-Hamiltonian and harmonic-coordinates --- have been shown to be
physically equivalent \cite{DJS01,ABF01}~: there exists a unique
``contact'' transformation of the binary's dynamical variables that
transforms the harmonic-coordinates Lagrangian \cite{ABF01} into a
different Lagrangian, whose Legendre transform agrees with the
ADM-coordinates Hamiltonian \cite{DJS00}.

In the works \cite{JaraS98,JaraS99,DJS00,DJS01} and
\cite{BF00,BFeom,BFreg,BFregM,ABF01} the compact objects are modelled
by structureless (non-spinning) point-particles. Such a modelling
is quite efficient and physically sound when describing the inspiral
of compact binaries, but the shortcoming is the necessity of a
regularization for removing the infinite self field of each of the
point-particles. The regularization of Hadamard (or, more precisely, a
refined form of it proposed in Refs. \cite{BFreg,BFregM} and
implemented in the harmonic-coordinates approach) has been applied but
turned out to be incomplete in the sense that one (and only one)
numerical coefficient remains undetermined at the 3PN order~:
$\omega_{\rm s}$ in the ADM-Hamiltonian formalism
\cite{JaraS98,JaraS99}, $\lambda$ in the harmonic-coordinates approach
\cite{BF00,BFeom}. This coefficient has been computed in
Ref. \cite{DJSdim} with the help of a dimensional regularization
instead of the Hadamard one, within the ADM-Hamiltonian formalism,
with the result $\omega_{\rm s}=0$ or equivalently
$\lambda=-\frac{1987}{3080}$ (below we shall keep the value of
$\lambda$ unspecified).

The present paper's lineage is the harmonic-coordinates approach
\cite{BF00,BFeom,BFreg,BFregM,ABF01}. Its goal is the completion of
the 3PN dynamics of compact binaries --- equations of motion,
Lagrangian, conserved integrals --- in the frame where the
center-of-mass is located at the origin of the coordinates. Our
motivation is that the center-of-mass equations of motion constitute
the needed starting point in many applications like the one in
Ref. \cite{MW02}. In Section II we recall the expression derived in
Ref. \cite{ABF01} for the position of the center-of-mass in an
arbitrary harmonic-coordinates frame. In Section III the individual
positions of the particles in the center-of-mass frame are obtained as
functions of the relative separation and velocity. We then compute the
3PN-accurate center-of-mass equations of motion. These equations are
substantially simpler than in a general frame --- though they are
still quite lengthy (that is unavoidable at such a high post-Newtonian
order). In particular we recover the center-of-mass equations of
motion at the 2.5PN order derived by Lincoln and Will \cite{LW90} on
the basis of the general-frame 2.5PN equations of Damour and Deruelle
\cite{DD81a,DD81b,D82}. In Section IV the 3PN relative Lagrangian (in
harmonic coordinates), describing the conservative part of the
dynamics, is obtained. Note that the center-of-mass relative
Lagrangian does not straightforwardly follow from the general-frame
Lagrangian of de Andrade, Blanchet and Faye \cite{ABF01}, because one
is not {\it a priori} allowed to use in a Lagrangian some expressions
which are consequences of the equations of motion derived from that
Lagrangian. We found it convenient to derive the center-of-mass
Lagrangian {\it ab initio} using some guess-work (i.e. adjusting a set
of coefficients in order to reproduce after Lagrangian variation the
correct equations of motion). From that center-of-mass Lagrangian we
then obtain in a standard way the Noetherian conserved energy and
angular momentum, thereby completing our harmonic-coordinates
approach.

Further investigations are proposed. Section V deals with the
connection between the center-of-mass Lagrangian and the
center-of-mass ADM-Hamiltonian. We check that the center-of-mass
reduction of the contact transformation worked out in
Ref. \cite{ABF01} between the harmonic-coordinates Lagrangian and the
ADM-coordinates Hamiltonian is identical --- as it must surely be ---
to the contact transformation connecting the center-of-mass versions
of these Lagrangian and Hamiltonian. In the process we recover the 3PN
Hamiltonian for the relative motion as computed by Jaranowski and
Sch\"afer \cite{JaraS98,JaraS99}. Finally in Section VI we consider
the problem of the stability against linear perturbations of circular
orbits. We undertake the problem by perturbing both the equations of
motion in harmonic coordinates, and the Hamiltonian equations in ADM
coordinates (the two methods give equivalent results). We obtain a
gauge-invariant criterion for the stability of circular orbits up to
the 3PN order.

\section{The center-of-mass vector position}

Our study starts with the expression, derived in Ref. \cite{ABF01},
for the position $G^i$ of the binary's center of mass. In this Section
we briefly review the construction of $G^i$. Note that the
center-of-mass position can also be interpreted as the gravitational
mass-type dipole moment. Actually, using a slight abuse of language,
by center-of-mass position $G^i$ we really mean the gravitational
dipole (its dimension is that of a mass times a length). In a future
work \cite{BImult} we shall show that the gravitational mass-type
dipole moment which follows from a 3PN wave-generation formalism
(instead of being inferred from the 3PN equations of motion) is in
complete agreement with the present center-of-mass vector $G^i$.

By equations of binary motion (in a given coordinate system) we mean
the acceleration $a_A^i(t)=dv_A^i/dt$ of body $A$, where $A=1,2$ and
the spatial index $i=1,2,3$, as a function of the positions $y_B^i$
and coordinate velocities $v_B^i(t)=dy_B^i/dt$. Eq. (7.16) in
Ref. \cite{BFeom} gives the 3PN equations of motion in the harmonic
coordinate system. The 3PN Lagrangian in harmonic coordinates
(considering only the conservative part of the dynamics) is given by
Eq. (4.1) in Ref. \cite{ABF01}; it takes the form

\begin{equation}\label{2}
L = L_{\rm N}[{\bf y}_A,{\bf v}_A]+\frac{1}{c^2}L_{\rm 1PN}[{\bf
y}_A,{\bf v}_A]+\frac{1}{c^4}L_{\rm 2PN}[{\bf y}_A,{\bf v}_A,{\bf
a}_A]+\frac{1}{c^6}L_{\rm 3PN}[{\bf y}_A,{\bf v}_A,{\bf a}_A]\;.
\end{equation}
The successive post-Newtonian orders depend on the positions and
velocities, and also, starting from the 2PN order, on the
accelerations. The fact that a harmonic-coordinates Lagrangian
necessarily becomes a ``generalized'' one (depending on accelerations)
at the 2PN approximation has been proved by Damour and Deruelle
\cite{DD81b}\footnote{This is consistent with a general argument of
Martin and Sanz \cite{MS79} that in coordinate systems which preserve
the Lorentz invariance (as the harmonic coordinates do) the equations
of motion at 2PN and higher orders cannot be derived from an ordinary
Lagrangian.}. At the 3PN order we found \cite{ABF01} that the
Lagrangian also depends on accelerations, but it is notable that these
accelerations are sufficient (i.e. there is no need to include
derivatives of accelerations). Furthermore the dependence upon the
accelerations at both the 2PN and 3PN orders is linear. Indeed one can
always eliminate from a generalized Lagrangian, taking the form of a
perturbative post-Newtonian expansion, a non-linear --- for instance
quadratic --- term in the accelerations by adding a ``double-zero''
counter term, whose Lagrangian variation is zero on-shell and
therefore which does not contribute to the dynamics (we refer to
\cite{DS85} for a general discussion on acceleration-dependent terms
in a post-Newtonian Lagrangian). The conservative part of the
equations of motion of body $A$ (neglecting the 2.5PN and 3.5PN
radiation damping terms) can be written with the help of the
variational derivative of the Lagrangian as

\begin{equation}\label{3}
\frac{\delta L}{\delta y_A^i} \equiv \frac{\partial L}{\partial y_A^i}
- \frac{d}{dt} \bigg( \frac{\partial L}{\partial v_A^i} \bigg)+
\frac{d^2}{dt^2} \bigg(\frac{\partial L}{\partial a_A^i} \bigg)={\cal
O}\left(\frac{1}{c^8}\right)\;.
\end{equation}
It is important to remember that here the equations of motion are
supposed to have been ``order-reduced'' using the equations themselves
at lower post-Newtonian order (i.e. any $a_A^i$ occuring in a
post-Newtonian term must be replaced by its expression in terms of the
$y_B^i$'s and $v_B^i$'s following the equations of motion).

The existence of a center-of-mass integral for the 3PN dynamics is the
consequence of its invariance under Lorentz transformations or
boosts. The Lorentz invariance of the (harmonic-coordinates) equations
of motion was established in Ref. \cite{BFeom}. Technically this means
that a specific variant, defined in \cite{BFregM}, of the Hadamard
regularization that respects the Lorentz invariance, is to be
implemented. Consider an infinitesimal deformation of the path of the
two particles, say $\delta y_A^i(t)\equiv {y'}_A^i(t)-y_A^i(t)$. Then
the corresponding perturbation of the Lagrangian, i.e. $\delta L =
L[{\bf y}'_A,{\bf v}'_A,{\bf a}'_A]-L[{\bf y}_A,{\bf v}_A,{\bf a}_A]$,
reads, to the linearized order,

\begin{equation}\label{5}
\delta L = \frac{d Q}{dt} + \sum_A \frac{\delta L}{\delta y_A^i}\delta
y_A^i+{\cal O}(\delta{\bf y}_A^2)\;.
\end{equation}
It involves the total time derivative of the function

\begin{equation}\label{6}
Q = \sum_A \left(p_A^i\delta y_A^i+q_A^i\delta v_A^i\right)\;,
\end{equation}
which is defined in terms of the momenta conjugate to the velocities
and accelerations,

\begin{subequations}\label{7}\begin{eqnarray}
p_A^i &=& \frac{\delta L}{\delta v_A^i}\equiv \frac{\partial
L}{\partial v_A^i}-\frac{d}{dt} \bigg(\frac{\partial L}{\partial
a_A^i} \bigg)\;, \label{7a}\\ q_A^i &=& \frac{\delta L}{\delta
a_A^i}\equiv \frac{\partial L}{\partial a_A^i}\;.\label{7b}
\end{eqnarray}\end{subequations}
Eqs. (\ref{5})-(\ref{7}) are nothing but the Noetherian equations in
the case of a generalized Lagrangian.

In the case of a Lorentz transformation, the change in the position of
particle $A$, to linear order in the (constant) boost velocity $V^i$,
is

\begin{equation}\label{8}
\delta y_A^i=-V^i t +\frac{1}{c^2}V^jy_A^j v_A^i+{\cal O}({\bf V}^2)\;.
\end{equation}
Because the 3PN dynamics is invariant under Lorentz boosts, the change
in the Lagrangian given by Eq. (\ref{5}) must take the form of a total
time derivative on-shell, i.e. when the equations of motion (\ref{3})
are satisfied. Hence there should exist a certain functional $Z^i$ of
the positions, velocities and accelerations such that (on-shell)

\begin{equation}\label{9}
\delta L=V^i \frac{dZ^i}{dt}+{\cal O}({\bf V}^2)\;.
\end{equation}
Using this, together with the particular form of the transformation
law (\ref{8}), into Eq. (\ref{5}) we readily obtain the conservation
(on-shell) of the Noetherian integral $K^i = G^i -t P^i$, where $P^i$,
the total linear momentum, and $G^i$, the center-of-mass position, are
given by

\begin{subequations}\label{10}\begin{eqnarray}
P^i &=& \sum_A p_A^i\;,\\
G^i &=& -Z^i+\sum_A\left(-q_A^i+\frac{1}{c^2}\left[y_A^ip_A^jv_A^j
+y_A^iq_A^ja_A^j+v_A^iq_A^jv_A^j\right]\right)\;.
\end{eqnarray}\end{subequations}
Since $P^i$ is itself constant [indeed, apply Eqs. (\ref{5})-(\ref{6})
to the case of a constant spatial translation~: $\delta
y_A^i=\epsilon^i$],

\begin{equation}\label{9'}
\frac{d P^i}{dt} = 0\;,
\end{equation}
we find that the conservation of $K^i$ implies that

\begin{equation}\label{11}
\frac{d G^i}{dt} = P^i\;.
\end{equation}
[We neglect terms of order ${\cal O}(c^{-8})$.] The center-of-mass
vector $G^i$ is conserved in the rest frame where $P^i=0$; it will be
zero, by definition, in the center-of-mass frame.
 
Applying these considerations to the 3PN equations of motion and
Lagrangian in harmonic coordinates, we found \cite{ABF01} that indeed
the variation of the Lagrangian uniquely defines some function $Z^i$
(this is a confirmation of the boost symmetry of the equations of
motion), and from it we explicitly determined the center-of-mass
vector position in an arbitrary harmonic-coordinates
frame\footnote{All-over this paper the gravitational constant is set
to $G=1$.}~:

\begin{eqnarray}\label{15} 
G^i &=& m_1 y_1^i \nonumber \\ &&+
\frac{m_1}{c^2}\Bigg\{\bigg(-\frac{m_2}{2 r_{12}} +
\frac{v_1^2}{2}\bigg) y_1^i\Bigg\} \nonumber \\ && +
\frac{m_1}{c^4}\Bigg\{m_2 \bigg(-\frac{7}{4} (n_{12}v_1) - \frac{7}{4}
(n_{12}v_2)\bigg) v_1^i \nonumber \\ & & \qquad\quad +
\Bigg[-\frac{5}{4}\frac{m_1 m_2}{r_{12}^2} +
\frac{7}{4}\frac{m_2^2}{r_{12}^2} + \frac{3 v_1^4}{8} \nonumber \\ & &
\qquad\quad\quad + \frac{m_2}{r_{12}} \bigg(-\frac{1}{8} (n_{12}v_1)^2
- \frac{1}{4} (n_{12}v_1) (n_{12}v_2) + \frac{1}{8} (n_{12}v_2)^2
\nonumber \\ & & \qquad\qquad\qquad\quad + \frac{19}{8} v_1^2 -
\frac{7}{4} (v_1v_2) - \frac{7}{8} v_2^2\bigg)\Bigg]y_1^i\Bigg\}
\nonumber \\ && + \frac{m_1}{c^6} \Bigg\{ \Bigg[
\frac{235}{24}\frac{m_1 m_2}{r_{12}}(n_{12}v_{12}) -
\frac{235}{24}\frac{m_2^2}{r_{12}} (n_{12}v_{12}) \nonumber \\ & &
\qquad\quad\quad + m_2 \bigg(\frac{5}{12} (n_{12}v_1)^3 + \frac{3}{8}
(n_{12}v_1)^2 (n_{12}v_2) + \frac{3}{8} (n_{12}v_1) (n_{12}v_2)^2
\nonumber \\ & & \qquad\qquad\qquad\quad + \frac{5}{12} (n_{12}v_2)^3
- \frac{15}{8} (n_{12}v_1) v_1^2 - (n_{12}v_2) v_1^2 + \frac{1}{4}
(n_{12}v_1) (v_1v_2) \nonumber \\ & & \qquad\qquad\qquad\quad +
\frac{1}{4} (n_{12}v_2) (v_1v_2) - (n_{12}v_1) v_2^2 - \frac{15}{8}
(n_{12}v_2) v_2^2\bigg) \Bigg] v_1^i \nonumber \\ & & \qquad\quad +
\Bigg[ \frac{5 v_1^6}{16} \nonumber \\ & & \qquad\quad\quad +
\frac{m_2}{r_{12}} \bigg(\frac{1}{16} (n_{12}v_1)^4 + \frac{1}{8}
(n_{12}v_1)^3 (n_{12}v_2) + \frac{3}{16} (n_{12}v_1)^2 (n_{12}v_2)^2
\nonumber \\ & & \qquad\qquad\qquad\quad + \frac{1}{4} (n_{12}v_1)
(n_{12}v_2)^3 - \frac{1}{16} (n_{12}v_2)^4 - \frac{5}{16}
(n_{12}v_1)^2 v_1^2 \nonumber \\ & & \qquad\qquad\qquad\quad -
\frac{1}{2} (n_{12}v_1) (n_{12}v_2) v_1^2 - \frac{11}{8} (n_{12}v_2)^2
v_1^2 + \frac{53}{16} v_1^4 + \frac{3}{8} (n_{12}v_1)^2 (v_1v_2)
\nonumber \\ & & \qquad\qquad\qquad\quad + \frac{3}{4} (n_{12}v_1)
(n_{12}v_2) (v_1v_2) + \frac{5}{4} (n_{12}v_2)^2 (v_1v_2) \nonumber \\
& & \qquad\qquad\qquad\quad - 5 v_1^2 (v_1v_2) + \frac{17}{8}
(v_1v_2)^2 - \frac{1}{4} (n_{12}v_1)^2 v_2^2 - \frac{5}{8} (n_{12}v_1)
(n_{12}v_2) v_2^2 \nonumber \\ & & \qquad\qquad\qquad\quad +
\frac{5}{16} (n_{12}v_2)^2 v_2^2 + \frac{31}{16} v_1^2 v_2^2 -
\frac{15}{8} (v_1v_2) v_2^2 - \frac{11}{16} v_2^4\bigg) \nonumber \\ &
& \qquad\quad\quad + \frac{m_1 m_2}{r_{12}^2} \bigg(\frac{79}{12}
(n_{12}v_1)^2 - \frac{17}{3} (n_{12}v_1) (n_{12}v_2) + \frac{17}{6}
(n_{12}v_2)^2 \nonumber \\ & & \qquad\qquad\qquad\quad -
\frac{175}{24} v_1^2 + \frac{40}{3} (v_1v_2) - \frac{20}{3}
v_2^2\bigg) \nonumber \\ & & \qquad\quad\quad + \frac{m_2^2}{r_{12}^2}
\bigg(-\frac{7}{3} (n_{12}v_1)^2 + \frac{29}{12} (n_{12}v_1)
(n_{12}v_2) + \frac{2}{3} (n_{12}v_2)^2 \nonumber \\ & &
\qquad\qquad\qquad\quad + \frac{101}{12} v_1^2 - \frac{40}{3} (v_1v_2)
+ \frac{139}{24} v_2^2 \bigg) \nonumber \\ & & \qquad\quad\quad
-\frac{19}{8}\frac{m_1 m_2^2}{r_{12}^3}\nonumber \\ & &
\qquad\quad\quad + \frac{m_1^2 m_2}{r_{12}^3} \bigg(\frac{13721}{1260}
- \frac{22}{3} \ln \Big(\frac{r_{12}}{r'_1} \Big)\bigg) \nonumber \\ &
& \qquad\quad\quad + \frac{m_2^3}{r_{12}^3} \bigg(-\frac{14351}{1260}
+ \frac{22}{3} \ln \Big(\frac{r_{12}}{r'_2} \Big)\bigg)\Bigg] y_1^i
\Bigg\} \nonumber\\&& + 1 \leftrightarrow 2 + {\cal O}\left(
\frac{1}{c^8} \right)\;.
\end{eqnarray}
To the terms given above we must add those corresponding to the
relabelling $1 \leftrightarrow 2$. We denote by $r_{12}=|{\bf
y}_1-{\bf y}_2|$, $n^i_{12}=(y^i_1-y^i_2)/r_{12}$ and
$v^i_{12}=v^i_1-v^i_2$ the relative particles' separation, unit
direction and velocity.

The expression (\ref{15}) has been systematically order-reduced using
the equations of motion and therefore depends only on the positions
and velocities (no accelerations). Notice the appearance at the 3PN
order of some logarithmic terms, containing two constants $r'_1$ and
$r'_2$ (one for each body) having the dimension of a length. It was
proved in Refs. \cite{BFeom,ABF01} that these logarithms, and the
$r'_A$'s therein, can be removed by an infinitesimal gauge
transformation at the 3PN order. Thus we can refer to the $r'_A$'s as
some ``gauge constants'', since they are merely associated with a
choice of coordinate system, and thereby do not carry any physical
meaning~: they will always cancel out when deriving some physical,
gauge-invariant, results. On the other hand we notice that $G^i$ is
free of the physical regularization ambiguity $\lambda$ present in the
equations of motion and Lagrangian.

The previous derivation of the center of mass neglected the effect of
radiation reaction. To take into account this effect we introduce some
appropriate modifications at the 2.5PN order of the linear momentum
and center of mass position~:

\begin{subequations}\label{15'}\begin{eqnarray}
{\widetilde P}^i&=& P^i+\left(\frac{4 m_1^2 m_2}{5 c^5 r_{12}^2}
n^i_{12}\left[v_{12}^2-\frac{2 m_1}{r_{12}}\right]+ 1 \leftrightarrow
2\right)\;,\label{15a'}\\ {\widetilde G}^i&=& G^i+\left(\frac{4 m_1
m_2}{5 c^5} v^i_1\left[v_{12}^2-\frac{2 (m_1+m_2)}{r_{12}}\right]+ 1
\leftrightarrow 2\right)\;.\label{15b'}
\end{eqnarray}\end{subequations}
With these definitions, we find that the conservation laws (\ref{9'})
and (\ref{11}), but now when taking into account the
radiation-reaction effect, take in fact exactly the same form~:

\begin{subequations}\label{15''}\begin{eqnarray}
\frac{d {\widetilde P}^i}{dt} &=& {\cal O}\left( \frac{1}{c^7}
\right)\;,\\ \frac{d {\widetilde G}^i}{dt} &=& {\widetilde P}^i + {\cal
O}\left( \frac{1}{c^7} \right)\;.
\end{eqnarray}\end{subequations}
This finding is quite normal~: recall that the total linear momentum
of an isolated system is conserved up to the 3PN order
included. Indeed the integral over the system of the local radiation
reaction forces is a total time derivative at the 2.5PN order, and
therefore it does not contribute to any change in the total linear
momentum. The modification of the linear momentum by radiation
reaction, or net radiation ``recoil'' of the source, is a smaller
effect, of order 3.5PN --- negligible in Eqs. (\ref{15''}).

\section{Equations of motion in the center-of-mass frame}

The positions and velocities of the two particles in the
center-of-mass frame at the 3PN order are obtained by solving the
equation

\begin{equation}\label{16}
{\widetilde G}^i[{\bf y}_A,{\bf v}_A] = {\cal O}\left(
\frac{1}{c^7}\right)\;,
\end{equation}
where ${\widetilde G}^i$ is defined in the previous Section. Obviously
the solution must be determined iteratively, in a post-Newtonian
perturbative sense, with systematic order-reduction of the
equations. At the Newtonian order we get

\begin{subequations}\label{16'}\begin{eqnarray}
y_1^i&=&X_2 x^i+{\cal O}\left( \frac{1}{c^2}\right)\;,\\ y_2^i&=&-X_1
x^i+{\cal O}\left( \frac{1}{c^2}\right)\;.
\end{eqnarray}\end{subequations}
In this paper we employ the following notation. The relative binary's
separation is

\begin{subequations}\label{19}\begin{eqnarray}
x^i&=&y_1^i-y_2^i\;,\\ r&=&|{\bf x}|\quad\hbox{and}\quad
n^i=\frac{x^i}{r}\;.
\end{eqnarray}\end{subequations}
For the relative velocity and acceleration we pose

\begin{subequations}\label{20}\begin{eqnarray}
v^i&=&\frac{dx^i}{dt}=v_1^i-v_2^i\quad\hbox{and} \quad\dot{r}={\bf
n}\cdot{\bf v}\;,\\ a^i&=&\frac{dv^i}{dt}\;,
\end{eqnarray}\end{subequations}
[$r$ and $v^i$ were formerly denoted $r_{12}$ and $v_{12}^i$ in
Eq. (\ref{15})]. Concerning the mass parameters we denote

\begin{subequations}\label{17}\begin{eqnarray}
X_1&=&\frac{m_1}{m}\quad\hbox{and}\quad
X_2=\frac{m_2}{m}\;,\\m&=&m_1+m_2\;,\\ \nu&=&\frac{m_1m_2}{m^2}=X_1
X_2\quad\hbox{and}\quad \mu=m\nu\;.
\end{eqnarray}\end{subequations}
All the expressions that are written in the center-of-mass frame are
conveniently parameterized by $m$ and the very useful mass ratio $\nu$
(such that $\nu=\frac{1}{4}$ in the equal-mass case and $\nu\to 0$ in
the test-mass limit for one of the bodies). Often it is convenient to
consider reduced quantities, i.e. quantities
divided by the reduced mass $\mu$.

The Newtonian solution (\ref{16'}) is inserted into the 1PN terms of
Eq. (\ref{16}) and we then obtain an equation for the 1PN corrections
in $y_1^i$ and $y_2^i$. Solving that equation we plug the result back
into the 1PN and 2PN terms of Eq. (\ref{16}) and obtain the 2PN
corrections in the same way. The process continues at the next order
and this finally results in the 3PN-accurate relationship between the
individual center-of-mass positions $y_1^i$ and $y_2^i$ and the
relative position $x^i$ and velocity $v^i$. In the course of the
computation we use for the order-reduction the center-of-mass
equations of relative motion at the 2PN order --- that is, at one
post-Newtonian order before the 3PN order we want to reach. Since we
give below the result for the 3PN equations, we do not detail this
step and simply present the final expressions. They are in the form

\begin{subequations}\label{21}\begin{eqnarray}
y_1^i&=&\Big[X_2+\nu (X_1-X_2) {\cal P}\Big] x^i +\nu (X_1-X_2){\cal
Q}\,v^i+ {\cal O}\left( \frac{1}{c^7} \right)\;,\label{21a}\\
y_2^i&=&\Big[-X_1+\nu (X_1-X_2) {\cal P}\Big] x^i +\nu (X_1-X_2) {\cal
Q}\,v^i+ {\cal O}\left( \frac{1}{c^7} \right)\;,\label{21b}
\end{eqnarray}\end{subequations}
where all the post-Newtonian corrections, beyond the Newtonian result
(\ref{16'}), are proportional to the mass ratio $\nu$ and the mass
difference $X_1-X_2$. The two dimensionless coefficients ${\cal P}$
and ${\cal Q}$ depend on the mass parameters $m$, $\nu$, the distance
$r$, the relative velocity $v^2={\bf v}^2$ and the radial velocity
$\dot{r}={\bf n}\cdot{\bf v}$~:

\begin{subequations}\label{22}\begin{eqnarray}
{\cal P}&=&\frac{1}{c^2}\bigg[\frac{v^2}{2}
-\frac{m}{2\,r}\bigg]\nonumber\\ &+& \frac{1}{c^4}\bigg[
\frac{3\,v^4}{8} - \frac{3\,\nu\,v^4}{2} \nonumber\\ && \qquad+
\frac{m}{r}\,\left( -\frac{\dot{r}^2}{8} + \frac{3\,\dot{r}^2\,\nu}{4}
+ \frac{19\,v^2}{8} + \frac{3\,\nu\,v^2}{2} \right)\nonumber\\ &&
\qquad+\frac{m^2}{r^2}\left(\frac{7}{4} - \frac{\nu}{2} \right)
\bigg]\nonumber\\ &+& \frac{1}{c^6}\bigg[\frac{5\,v^6}{16} -
\frac{11\,\nu\,v^6}{4} + 6\,\nu^2\,v^6 \nonumber\\ && \qquad
+\frac{m}{r}\left( \frac{\dot{r}^4}{16} - \frac{5\,\dot{r}^4\,\nu}{8}
+ \frac{21\,\dot{r}^4\,\nu^2}{16} - \frac{5\,\dot{r}^2\,v^2}{16} +
\frac{21\,\dot{r}^2\,\nu\,v^2}{16} \right.\nonumber\\ &&
\qquad\qquad\qquad -\left.  \frac{11\,\dot{r}^2\,\nu^2\,v^2}{2} +
\frac{53\,v^4}{16} - 7\,\nu\,v^4 - \frac{15\,\nu^2\,v^4}{2} \right)
\nonumber\\ && \qquad +\frac{m^2}{r^2}\left( -\frac{7\,\dot{r}^2}{3} +
\frac{73\,\dot{r}^2\,\nu}{8} + 4\,\dot{r}^2\,\nu^2 +
\frac{101\,v^2}{12} - \frac{33\,\nu\,v^2}{8} + 3\,\nu^2\,v^2 \right)
\nonumber\\ && \qquad + \frac{m^3}{r^3}\left( -\frac{14351}{1260} +
\frac{\nu}{8} - \frac{\nu^2}{2} + \frac{22}{3}\,\ln
\Big(\frac{r}{r''_0}\Big) \right) \bigg]\;,\\ {\cal
Q}&=&\frac{1}{c^4}\bigg[-\frac{7\,m\,\dot{r}}{4}\bigg] \nonumber\\&+&
\frac{1}{c^5}\bigg[\frac{4\,m\,v^2}{5} -\frac{8\,m^2}{5\,r}
\bigg]\nonumber\\ &+& \frac{1}{c^6}\bigg[ m\,\dot{r}\left(
\frac{5\,\dot{r}^2}{12} - \frac{19\,\dot{r}^2\,\nu}{24} -
\frac{15\,v^2}{8} + \frac{21\,\nu\,v^2}{4} \right) \nonumber\\ &&
\qquad + \frac{m^2\,\dot{r}}{r}\left( -\frac{235}{24}-
\frac{21\,\nu}{4} \right)\bigg]\;.
\end{eqnarray}\end{subequations}
Up to the 2.5PN order we find agreement with the circular-orbit limit
of Eqs. (6.4) in Ref. \cite{B96} (notice that the 2.5PN
radiation-reaction term itself is proportional to the velocity and so
it enters only the coefficient ${\cal Q}$).

In Eq. (\ref{22}) we find that the logarithms appear at the 3PN order
and only in the coefficient ${\cal P}$. They contain a particular
combination $r''_0$ of the two gauge-constants $r'_1$ and $r'_2$ that
is defined by

\begin{equation}\label{23}
(X_1-X_2) \ln r''_0 = X_1^2 \ln r'_1 - X_2^2 \ln r'_2\;.
\end{equation}
This constant $r''_0$ happens to be different from a similar constant
$r'_0$ which will have to be introduced to the 3PN equations of
relative motion and Lagrangian [see Eq. (\ref{26}) below].

The 3PN center-of-mass equations of motion are obtained in a
straightforward way by replacing in the general-frame 3PN equations
derived in Ref. \cite{BFeom} (see Eq. (7.16) there) the positions by
Eqs. (\ref{21})-(\ref{22}), and the velocities by the derivatives of
Eqs. (\ref{21})-(\ref{22}) (applying as usual the order-reduction of
all accelerations where necessary). Actually for this purpose we do
not need the Eqs. (\ref{21})-(\ref{22}) with the full 3PN precision;
the 2PN-accurate ones are sufficient. We write the relative
acceleration in the center-of-mass frame in the form

\begin{equation}\label{24}
\frac{d v^i}{dt}=-\frac{m}{r^2}\Big[(1+{\cal A})\,n^i + {\cal B}\,v^i
\Big]+ {\cal O}\left( \frac{1}{c^7} \right)\;,
\end{equation}
and find that the coefficients ${\cal A}$ and ${\cal B}$ are

\begin{subequations}\label{25}\begin{eqnarray}
{\cal A}&=& \frac{1}{c^2}\left\{-\frac{3\,\dot{r}^2\,\nu}{2} + v^2 +
3\,\nu\,v^2-\frac{m}{r}\left(4 +2\,\nu \right)\right\}\nonumber\\ &+&
\frac{1}{c^4}\left\{\frac{15\,\dot{r}^4\,\nu}{8} -
\frac{45\,\dot{r}^4\,\nu^2}{8} - \frac{9\,\dot{r}^2\,\nu\,v^2}{2} +
6\,\dot{r}^2\,\nu^2\,v^2 + 3\,\nu\,v^4 -
4\,\nu^2\,v^4\right. \nonumber\\ &&\qquad + \left.\frac{m}{r}\left(
-2\,\dot{r}^2 - 25\,\dot{r}^2\,\nu - 2\,\dot{r}^2\,\nu^2 -
\frac{13\,\nu\,v^2}{2} + 2\,\nu^2\,v^2 \right)\right. \nonumber\\
&&\qquad +\left.\frac{m^2}{r^2}\,\left( 9 + \frac{87\,\nu}{4}
\right)\right\}\nonumber\\&+&\frac{1}{c^5}\left\{-
\frac{24\,\dot{r}\,\nu\,v^2}{5}\frac{m}{r}-\frac{136\,\dot{r}\,\nu}{15}
\frac{m^2}{r^2}\right\}\nonumber\\ &+&
\frac{1}{c^6}\left\{-\frac{35\,\dot{r}^6\,\nu}{16} +
\frac{175\,\dot{r}^6\,\nu^2}{16} -
\frac{175\,\dot{r}^6\,\nu^3}{16}+\frac{15\,\dot{r}^4\,\nu\,v^2}{2}
\right.\nonumber\\&&\qquad - \left.
\frac{135\,\dot{r}^4\,\nu^2\,v^2}{4} +
\frac{255\,\dot{r}^4\,\nu^3\,v^2}{8} -
\frac{15\,\dot{r}^2\,\nu\,v^4}{2} +
\frac{237\,\dot{r}^2\,\nu^2\,v^4}{8} \right.\nonumber\\ &&\qquad
-\left. \frac{45\,\dot{r}^2\,\nu^3\,v^4}{2} + \frac{11\,\nu\,v^6}{4} -
\frac{49\,\nu^2\,v^6}{4} + 13\,\nu^3\,v^6 \right.\nonumber\\ &&\qquad
+ \left.\frac{m}{r}\left( 79\,\dot{r}^4\,\nu -
\frac{69\,\dot{r}^4\,\nu^2}{2} - 30\,\dot{r}^4\,\nu^3 -
121\,\dot{r}^2\,\nu\,v^2 + 16\,\dot{r}^2\,\nu^2\,v^2
\right.\right.\nonumber\\&&\qquad\qquad\quad~ +\left.\left.
20\,\dot{r}^2\,\nu^3\,v^2+\frac{75\,\nu\,v^4}{4} + 8\,\nu^2\,v^4 -
10\,\nu^3\,v^4 \right)\right.\nonumber\\ &&\qquad + \left.
\frac{m^2}{r^2}\,\left( \dot{r}^2 + \frac{32573\,\dot{r}^2\,\nu}{168}
+ \frac{11\,\dot{r}^2\,\nu^2}{8} - 7\,\dot{r}^2\,\nu^3 +
\frac{615\,\dot{r}^2\,\nu\,\pi^2}{64} - \frac{26987\,\nu\,v^2}{840}
\right.\right.\nonumber\\&&\qquad\qquad\quad~ +\left.\left.
\nu^3\,v^2 - \frac{123\,\nu\,\pi^2\,v^2}{64} -
110\,\dot{r}^2\,\nu\,\ln \Big(\frac{r}{r'_0}\Big) + 22\,\nu\,v^2\,\ln
\Big(\frac{r}{r'_0}\Big) \right)\right.\nonumber\\&&\qquad
+\left.\frac{m^3}{r^3}\left( -16 - \frac{41911\,\nu}{420} +
\frac{44\,\lambda\,\nu}{3} - \frac{71\,\nu^2}{2} + \frac{41\,\nu\,{\pi
}^2}{16} \right)\right\}\;,\label{25a}\\ {\cal
B}&=&\frac{1}{c^2}\Big\{-4\,\dot{r} + 2\,\dot{r}\,\nu\Big\}
\nonumber\\&+&\frac{1}{c^4}\left\{\frac{9\,\dot{r}^3\,\nu}{2} +
3\,\dot{r}^3\,\nu^2 -\frac{15\,\dot{r}\,\nu\,v^2}{2} -
2\,\dot{r}\,\nu^2\,v^2\right.\nonumber\\ &&\qquad +
\left.\frac{m}{r}\left( 2\,\dot{r} + \frac{41\,\dot{r}\,\nu}{2} +
4\,\dot{r}\,\nu^2 \right)\right\}\nonumber\\&+&\frac{1}{c^5}\left\{
\frac{8\,\nu\,v^2}{5}\frac{m}{r}+\frac{24\,\nu}{5}
\frac{m^2}{r^2}\right\}\nonumber\\ &+&
\frac{1}{c^6}\left\{-\frac{45\,\dot{r}^5\,\nu}{8} +
15\,\dot{r}^5\,\nu^2 + \frac{15\,\dot{r}^5\,\nu^3}{4} +
12\,\dot{r}^3\,\nu\,v^2 \right.\nonumber\\&&\qquad
-\left. \frac{111\,\dot{r}^3\,\nu^2\,v^2}{4}
-12\,\dot{r}^3\,\nu^3\,v^2 -\frac{65\,\dot{r}\,\nu\,v^4}{8} +
19\,\dot{r}\,\nu^2\,v^4 + 6\,\dot{r}\,\nu^3\,v^4
\right.\nonumber\\&&\qquad\left. + \frac{m}{r}\left(
\frac{329\,\dot{r}^3\,\nu}{6} + \frac{59\,\dot{r}^3\,\nu^2}{2} +
18\,\dot{r}^3\,\nu^3 - 15\,\dot{r}\,\nu\,v^2 - 27\,\dot{r}\,\nu^2\,v^2
- 10\,\dot{r}\,\nu^3\,v^2 \right) \right.\nonumber\\&&\qquad
+\left.\frac{m^2}{r^2}\,\left( -4\,\dot{r} -
\frac{18169\,\dot{r}\,\nu}{840} + 25\,\dot{r}\,\nu^2 +
8\,\dot{r}\,\nu^3 - \frac{123\,\dot{r}\,\nu\,\pi^2}{32}
\right.\right.\nonumber\\&&\qquad\qquad\quad~\left.\left. +
44\,\dot{r}\,\nu\,\ln \Big(\frac{r}{r'_0}\Big)
\right)\right\}\;.\label{25b}
\end{eqnarray}\end{subequations}
Up to the 2.5PN order the result agrees with the one given by Lincoln
and Will \cite{LW90}. At the 3PN order we have some gauge-dependent
logarithms containing a constant $r'_0$ --- distinct from $r''_0$
introduced in Eq. (\ref{23})\footnote{They are
related by
$$(X_1-X_2)\ln\left(\frac{r''_0}{r'_0}\right) = X_1X_2
\ln\left(\frac{r'_1}{r'_2}\right)\;.
$$
} --- which is the ``logarithmic barycenter'' of the two constants
$r'_1$ and $r'_2$~:

\begin{equation}\label{26}
\ln r'_0 = X_1 \ln r'_1 + X_2 \ln r'_2\;.
\end{equation}
In addition there is the physical ambiguity $\lambda$ due to the
Hadamard self-field regularization ($\lambda$ cannot be removed by any
coordinate transformation); it appears at the 3PN order in the ${\cal
A}$-coefficient.

\section{Lagrangian and Noetherian conserved integrals}

The Lagrangian for the relative center-of-mass motion is obtained from
the 3PN center-of-mass equations of motion (\ref{24})-(\ref{25}) in
which one ignores for a moment the radiation-reaction 2.5PN term. The
Lagrangian in harmonic coordinates will necessarily be a generalized
one, depending on accelerations, from the 2PN order. At the 3PN order,
further acceleration terms are necessary but we do not need to include
derivatives of accelerations. Furthermore we can always restrict
ourselves to a Lagrangian that is linear in the accelerations. Hence,
our center-of-mass Lagrangian, denoted with a calligraphic letter
${\cal L}$ to distinguish it from the general-frame Lagrangian $L$, is
of the form

\begin{equation}\label{27}
{\cal L}={\cal L}_{\rm N}[{\bf x},{\bf v}]+\frac{1}{c^2}{\cal L}_{\rm
1PN}[{\bf x},{\bf v}]+\frac{1}{c^4}{\cal L}_{\rm 2PN}[{\bf x},{\bf
v},{\bf a}]+\frac{1}{c^6}{\cal L}_{\rm 3PN}[{\bf x},{\bf v},{\bf a}]\;.
\end{equation}

We recall that there is a large freedom for choosing a Lagrangian
because we can always add to it the total time derivative of an
arbitrary function. As a matter of convenience, we shall choose below
a particular center-of-mass Lagrangian in such a way that it is
``close'' (in the sense that many coefficients are identical) to some
``fictitious'' Lagrangian that is obtained from the general-frame one
given in Ref. \cite{ABF01} by the mere Newtonian replacements
$y_1^i\rightarrow X_2 x^i$, $y_2^i\rightarrow -X_1 x^i$. We
immediately point out that such a fictitious Lagrangian is {\it not}
the correct Lagrangian for describing the center-of-mass relative
motion. Indeed, the actual center-of-mass variables involve many
post-Newtonian corrections given by Eqs. (\ref{21})-(\ref{22}), so the
actual center-of-mass Lagrangian must contain some extra terms in
addition to those of the previous fictitious Lagrangian. However, we
find that these extra terms arise only at the 2PN order and not
before. We did not try to find a general method for obtaining
systematically the center-of-mass Lagrangian given the general-frame
one. Though such a method might exist it was in fact simpler to
proceed by guess-work, i.e. to introduce some unknown coefficients in
front of all possible types of terms, and to adjust these coefficients
so that the Lagrangian reproduces the correct equations of motion. Our
result (when divided by the reduced mass $\mu=m\nu$) is then

\begin{eqnarray}\label{27'}
\frac{{\cal L}}{\mu}&=& \frac{v^2}{2} + \frac{m}{r} \nonumber\\ &&
+\frac{1}{c^2}\bigg\{\frac{v^4}{8} - \frac{3\,\nu\,v^4}{8} +
\frac{m}{r}\,\left( \frac{\dot{r}^2\,\nu}{2} + \frac{3\,v^2}{2} +
\frac{\nu\,v^2}{2} \right)-\frac{m^2}{2\,r^2} \bigg\}\nonumber\\ &&
+\frac{1}{c^4}\bigg\{ \frac{v^6}{16} - \frac{7\,\nu\,v^6}{16} +
\frac{13\,\nu^2\,v^6}{16} \nonumber\\ &&\qquad~ + \frac{m}{r}\,\left(
\frac{3\,\dot{r}^4\,\nu^2}{8} - \frac{\dot{r}^2\,a_n\,\nu\,r}{8} +
\frac{\dot{r}^2\,\nu\,v^2}{4} - \frac{5\,\dot{r}^2\,\nu^2\,v^2}{4} +
\frac{7\,a_n\,\nu\,r\,v^2}{8} \right.\nonumber\\ &&\qquad\qquad\quad~
+ \left.\frac{7\,v^4}{8} - \frac{5\,\nu\,v^4}{4} -
\frac{9\,\nu^2\,v^4}{8} - \frac{7\,\dot{r}\,\nu\,r\,a_v}{4} \right)
\nonumber\\ &&\qquad~ +\frac{m^2}{r^2}\,\left( \frac{\dot{r}^2}{2} +
\frac{41\,\dot{r}^2\,\nu}{8} + \frac{3\,\dot{r}^2\,\nu^2}{2} +
\frac{7\,v^2}{4} - \frac{27\,\nu\,v^2}{8} + \frac{\nu^2\,v^2}{2}
\right) \nonumber\\ &&\qquad~ +\frac{m^3}{r^3}\,\left( \frac{1}{2} +
\frac{15\,\nu}{4} \right)\bigg\}\nonumber\\ &&
+\frac{1}{c^6}\bigg\{\frac{5\,v^8}{128} - \frac{59\,\nu\,v^8}{128} +
\frac{119\,\nu^2\,v^8}{64} - \frac{323\,\nu^3\,v^8}{128} \nonumber\\
&&\qquad~ + \frac{m}{r}\,\left( \frac{5\,\dot{r}^6\,\nu^3}{16} +
\frac{\dot{r}^4\,a_n\,\nu\,r}{16} -
\frac{5\,\dot{r}^4\,a_n\,\nu^2\,r}{16} -
\frac{3\,\dot{r}^4\,\nu\,v^2}{16} \right.\nonumber\\
&&\qquad\qquad\quad~\left.+ \frac{7\,\dot{r}^4\,\nu^2\,v^2}{4} -
\frac{33\,\dot{r}^4\,\nu^3\,v^2}{16} -
\frac{3\,\dot{r}^2\,a_n\,\nu\,r\,v^2}{16} -
\frac{\dot{r}^2\,a_n\,\nu^2\,r\,v^2}{16} \right.\nonumber\\
&&\qquad\qquad\quad~\left.+ \frac{5\,\dot{r}^2\,\nu\,v^4}{8} -
3\,\dot{r}^2\,\nu^2\,v^4 +\frac{75\,\dot{r}^2\,\nu^3\,v^4}{16} +
\frac{7\,a_n\,\nu\,r\,v^4}{8} \right.\nonumber\\
&&\qquad\qquad\quad~\left.- \frac{7\,a_n\,\nu^2\,r\,v^4}{2} +
\frac{11\,v^6}{16} - \frac{55\,\nu\,v^6}{16} + \frac{5\,\nu^2\,v^6}{2}
\right.\nonumber\\ &&\qquad\qquad\quad~ +\left.
\frac{65\,\nu^3\,v^6}{16} + \frac{5\,\dot{r}^3\,\nu\,r\,a_v}{12} -
\frac{13\,\dot{r}^3\,\nu^2\,r\,a_v}{8} \right.\nonumber\\
&&\qquad\qquad\quad~\left.- \frac{37\,\dot{r}\,\nu\,r\,v^2\,a_v}{8} +
\frac{35\,\dot{r}\,\nu^2\,r\,v^2\,a_v}{4} \right) \nonumber\\
&&\qquad~ + \frac{m^2}{r^2}\,\left( -\frac{109\,\dot{r}^4\,\nu}{144} -
\frac{259\,\dot{r}^4\,\nu^2}{36} + 2\,\dot{r}^4\,\nu^3 -
\frac{17\,\dot{r}^2\,a_n\,\nu\,r}{6} \right.\nonumber\\
&&\qquad\qquad\quad~ +\left.  \frac{97\,\dot{r}^2\,a_n\,\nu^2\,r}{12}
+\frac{\dot{r}^2\,v^2}{4} - \frac{41\,\dot{r}^2\,\nu\,v^2}{6} -
\frac{2287\,\dot{r}^2\,\nu^2\,v^2}{48} \right.\nonumber\\
&&\qquad\qquad\quad~ -\left.  \frac{27\,\dot{r}^2\,\nu^3\,v^2}{4} +
\frac{203\,a_n\,\nu\,r\,v^2}{12} + \frac{149\,a_n\,\nu^2\,r\,v^2}{6}
\right.\nonumber\\ &&\qquad\qquad\quad~ +\left. \frac{45\,v^4}{16} +
\frac{53\,\nu\,v^4}{24} + \frac{617\,\nu^2\,v^4}{24} -
\frac{9\,\nu^3\,v^4}{4} \right.\nonumber\\ &&\qquad\qquad\quad~
-\left. \frac{235\,\dot{r}\,\nu\,r\,a_v}{24} +
\frac{235\,\dot{r}\,\nu^2\,r\,a_v}{6} \right) \nonumber\\ &&\qquad~ +
\frac{m^3}{r^3}\,\left( \frac{3\,\dot{r}^2}{2} -
\frac{12041\,\dot{r}^2\,\nu}{420} + \frac{37\,\dot{r}^2\,\nu^2}{4} +
\frac{7\,\dot{r}^2\,\nu^3}{2} - \frac{123\,\dot{r}^2\,\nu\,{\pi
}^2}{64} \right.\nonumber\\ &&\qquad\qquad\quad~
+\left. \frac{5\,v^2}{4} + \frac{387\,\nu\,v^2}{70} -
\frac{7\,\nu^2\,v^2}{4} + \frac{\nu^3\,v^2}{2} + \frac{41\,\nu\,{\pi
}^2\,v^2}{64} \right.\nonumber\\ &&\qquad\qquad\quad~ \left. +
22\,\dot{r}^2\,\nu\,\ln \Big(\frac{r}{r'_0}\Big) -
\frac{22\,\nu\,v^2}{3}\ln \Big(\frac{r}{r'_0}\Big) \right)\nonumber\\
&&\qquad~ + \frac{m^4}{r^4}\,\left( -\frac{3}{8}-
\frac{2747\,\nu}{140} + \frac{11\,\lambda\,\nu}{3} +
\frac{22\,\nu}{3}\ln \Big(\frac{r}{r'_0}\Big) \right) \bigg\}\;.
\end{eqnarray}
Witness the acceleration terms present at the 2PN and 3PN orders~: our
notation is $a_n\equiv {\bf a}\cdot{\bf n}$ and $a_v\equiv {\bf a}\cdot{\bf
v}$ for the scalar products between $a^i=dv^i/dt$ and the direction
$n^i$ and velocity $v^i$. The conservative part of the equations of
motion is then identical (after order-reduction of the accelerations)
to

\begin{equation}\label{28}
\frac{\delta {\cal L}}{\delta x^i}\equiv \frac{\partial {\cal
L}}{\partial x^i} -\frac{d}{dt}\left(\frac{\partial {\cal L}}{\partial
v^i}\right) +\frac{d^2}{dt^2}\left(\frac{\partial {\cal L}}{\partial
a^i}\right)={\cal O}\left( \frac{1}{c^8} \right)\;.
\end{equation}

From the Lagrangian one deduces the conserved energy and angular
momentum using the generalized formulas [neglecting ${\cal
O}(c^{-8})$]

\begin{subequations}\label{28'}\begin{eqnarray}
E &=& v^i p^i + a^i q^i - {\cal L}\;,\\
J^i &=& \varepsilon_{ijk} \Big(x^j p^k + v^j q^k\Big)\;,
\end{eqnarray}\end{subequations}
(the first one being a generalized version of the Legendre transform),
where the conjugate momenta read

\begin{subequations}\label{28''}\begin{eqnarray}
p^i &=& \frac{\delta {\cal L}}{\delta v^i}\equiv \frac{\partial {\cal
L}}{\partial v^i}-\frac{d}{dt} \bigg(\frac{\partial {\cal L}}{\partial
a^i} \bigg)\;, \label{28a''}\\ q^i &=& \frac{\delta {\cal L}}{\delta
a^i}\equiv \frac{\partial {\cal L}}{\partial a^i}\;.\label{28b''}
\end{eqnarray}\end{subequations}
Alternatively one can compute the center-of-mass energy and angular
momentum directly from the general-frame quantities $E$ and $J^i$
given by Eqs. (4.2) and (4.4) in Ref. \cite{ABF01} by replacing all
variables by their center-of-mass expressions given by
Eqs. (\ref{21})-(\ref{22}); the result is the same. For the energy we
obtain

\begin{eqnarray}\label{30}
\frac{E}{\mu}&=& \frac{v^2}{2}-\frac{m}{r} \nonumber\\ &&
+\frac{1}{c^2}\bigg\{\frac{3\,v^4}{8} - \frac{9\,\nu\,v^4}{8} +
\frac{m}{r}\,\left( \frac{\dot{r}^2\,\nu}{2} + \frac{3\,v^2}{2} +
\frac{\nu\,v^2}{2} \right)+\frac{m^2}{2r^2}\bigg\}\nonumber\\ &&
+\frac{1}{c^4}\bigg\{\frac{5\,v^6}{16} - \frac{35\,\nu\,v^6}{16} +
\frac{65\,\nu^2\,v^6}{16} \nonumber\\ &&\qquad~ + \frac{m}{r}\,\left(
-\frac{3\,\dot{r}^4\,\nu}{8} + \frac{9\,\dot{r}^4\,\nu^2}{8} +
\frac{\dot{r}^2\,\nu\,v^2}{4} - \frac{15\,\dot{r}^2\,\nu^2\,v^2}{4} +
\frac{21\,v^4}{8} - \frac{23\,\nu\,v^4}{8} - \frac{27\,\nu^2\,v^4}{8}
\right)\nonumber\\ &&\qquad~ +\frac{m^2}{r^2}\,\left(
\frac{\dot{r}^2}{2} + \frac{69\,\dot{r}^2\,\nu}{8} +
\frac{3\,\dot{r}^2\,\nu^2}{2} + \frac{7\,v^2}{4} -
\frac{55\,\nu\,v^2}{8} + \frac{\nu^2\,v^2}{2} \right) \nonumber\\
&&\qquad~+\frac{m^3}{r^3}\,\left( -\frac{1}{2}- \frac{15\,\nu}{4}
\right)\bigg\}\nonumber\\ &&
+\frac{1}{c^6}\bigg\{\frac{35\,v^8}{128} - \frac{413\,\nu\,v^8}{128} +
\frac{833\,\nu^2\,v^8}{64} - \frac{2261\,\nu^3\,v^8}{128} \nonumber\\
&&\qquad~ + \frac{m}{r}\,\left( \frac{5\,\dot{r}^6\,\nu}{16} -
\frac{25\,\dot{r}^6\,\nu^2}{16} + \frac{25\,\dot{r}^6\,\nu^3}{16} -
\frac{9\,\dot{r}^4\,\nu\,v^2}{16} +
\frac{21\,\dot{r}^4\,\nu^2\,v^2}{4} \right.\nonumber\\
&&\qquad\qquad\quad~ \left. -\frac{165\,\dot{r}^4\,\nu^3\,v^2}{16} -
\frac{21\,\dot{r}^2\,\nu\,v^4}{16} -
\frac{75\,\dot{r}^2\,\nu^2\,v^4}{16} +
\frac{375\,\dot{r}^2\,\nu^3\,v^4}{16} \right.\nonumber\\
&&\qquad\qquad\quad~ \left. + \frac{55\,v^6}{16} -
\frac{215\,\nu\,v^6}{16} + \frac{29\,\nu^2\,v^6}{4} +
\frac{325\,\nu^3\,v^6}{16} \right) \nonumber\\ &&\qquad~ +
\frac{m^2}{r^2}\,\left( -\frac{731\,\dot{r}^4\,\nu}{48} +
\frac{41\,\dot{r}^4\,\nu^2}{4} + 6\,\dot{r}^4\,\nu^3 +
\frac{3\,\dot{r}^2\,v^2}{4} + \frac{31\,\dot{r}^2\,\nu\,v^2}{2}
\right.\nonumber\\ &&\qquad\qquad\quad~
\left. -\frac{815\,\dot{r}^2\,\nu^2\,v^2}{16} -
\frac{81\,\dot{r}^2\,\nu^3\,v^2}{4} + \frac{135\,v^4}{16} -
\frac{97\,\nu\,v^4}{8} + \frac{203\,\nu^2\,v^4}{8} -
\frac{27\,\nu^3\,v^4}{4} \right) \nonumber\\ &&\qquad~ +
\frac{m^3}{r^3}\,\left( \frac{3\,\dot{r}^2}{2} +
\frac{803\,\dot{r}^2\,\nu}{840} + \frac{51\,\dot{r}^2\,\nu^2}{4} +
\frac{7\,\dot{r}^2\,\nu^3}{2} - \frac{123\,\dot{r}^2\,\nu\,{\pi
}^2}{64} + \frac{5\,v^2}{4} \right.\nonumber\\ &&\qquad\qquad\quad~
\left. -\frac{6747\,\nu\,v^2}{280} - \frac{21\,\nu^2\,v^2}{4} +
\frac{\nu^3\,v^2}{2} + \frac{41\,\nu\,\pi^2\,v^2}{64}
\right.\nonumber\\ &&\qquad\qquad\quad~ \left. +
22\,\dot{r}^2\,\nu\,\ln \Big(\frac{r}{r'_0}\Big) -
\frac{22\,\nu\,v^2\,}{3}\ln \Big(\frac{r}{r'_0}\Big)
\right)\nonumber\\ &&\qquad~ + \frac{m^4}{r^4}\,\left( \frac{3}{8} +
\frac{2747\,\nu}{140} - \frac{11\,\lambda\,\nu}{3} -
\frac{22\,\nu}{3}\,\ln \Big(\frac{r}{r'_0}\Big) \right) \bigg\}\;.
\end{eqnarray}
As for the center-of-mass angular momentum we get

\begin{eqnarray}\label{31}
\frac{J^i}{\mu} &=& \varepsilon_{ijk}x^jv^k \Bigg[ 1 \nonumber\\ &&
+\frac{1}{c^2} \bigg\{\left(1 - 3\, \nu \right)\frac{v^2}{2} +
\frac{m}{r}\,\left( 3 + \nu \right)\bigg\}\nonumber\\ &&
+\frac{1}{c^4}\bigg\{ \frac{3\,v^4}{8} - \frac{21\,\nu\,v^4}{8} +
\frac{39\,\nu^2\,v^4}{8} \nonumber\\ &&\qquad~ + \frac{m}{r}\,\left(
-\dot{r}^2\,\nu - \frac{5\,\dot{r}^2\,\nu^2}{2} + \frac{7\,v^2}{2} -
5\,\nu\,v^2 - \frac{9\,\nu^2\,v^2}{2} \right)\nonumber\\ &&\qquad~
+\frac{m^2}{r^2}\,\left( \frac{7}{2} - \frac{41\,\nu}{4} + \nu^2
\right)\bigg\}\nonumber\\ && +\frac{1}{c^6}\bigg\{\frac{5\,v^6}{16} -
\frac{59\,\nu\,v^6}{16} + \frac{119\,\nu^2\,v^6}{8} -
\frac{323\,\nu^3\,v^6}{16} \nonumber\\ &&\qquad~ + \frac{m}{r}\,\left(
\frac{3\,\dot{r}^4\,\nu}{4} - \frac{3\,\dot{r}^4\,\nu^2}{4}
-\frac{33\,\dot{r}^4\,\nu^3}{8} - 3\,\dot{r}^2\,\nu\,v^2 +
\frac{7\,\dot{r}^2\,\nu^2\,v^2}{4} \right.\nonumber\\
&&\qquad\qquad\quad~ +\left.  \frac{75\,\dot{r}^2\,\nu^3\,v^2}{4} +
\frac{33\,v^4}{8} - \frac{71\,\nu\,v^4}{4} + \frac{53\,\nu^2\,v^4}{4}
+ \frac{195\,\nu^3\,v^4}{8} \right) \nonumber\\ &&\qquad~ +
\frac{m^2}{r^2}\,\left( \frac{\dot{r}^2}{2} -
\frac{287\,\dot{r}^2\,\nu}{24} - \frac{317\,\dot{r}^2\,\nu^2}{8} -
\frac{27\,\dot{r}^2\,\nu^3}{2} + \frac{45\,v^2}{4} \right.\nonumber\\
&&\qquad\qquad\quad~ -\left.  \frac{161\,\nu\,v^2}{6} +
\frac{105\,\nu^2\,v^2}{4} - 9\,\nu^3\,v^2 \right) \nonumber\\
&&\qquad~ + \frac{m^3}{r^3}\,\left( \frac{5}{2} -
\frac{5199\,\nu}{280} - 7\,\nu^2 + \nu^3 + \frac{41\,\nu\,{\pi
}^2}{32} - \frac{44\,\nu}{3}\ln \Big(\frac{r}{r'_0}\Big)
\right)\bigg\}\Bigg]\;.
\end{eqnarray}
(The energy involves the regularization-ambiguity $\lambda$, while the
angular momentum is free of any physical ambiguity.) These quantities
are conserved in the sense that their time variation equals the
radiation-reaction effect. 
One can therefore modify them with terms of ``odd order'' to take into
account the radiation reaction due to gravitational wave emission. For
instance, in the leading 2.5PN radiation reaction one conventionally
chooses that the right-hand side of the balance equations for energy
and angular momentum take the standard form appropriate to the
quadrupolar approximation. We then pose,

\begin{subequations}\label{29'}\begin{eqnarray}
{\widetilde E}&=&
E+\frac{8\,m^3\,\dot{r}\,\nu^2}{5\,c^5\,r^2}v^2\;,\label{29a'}\\
{\widetilde J}^i&=& J^i-\frac{8\,m^3\,\dot{r}\,\nu^2}{5\,c^5\,r^2}
\varepsilon_{ijk}x^jv^k\;.\label{29b'}
\end{eqnarray}\end{subequations}
This choice is in agreement with the results of
\cite{IW93,IW95,GII97}.  Then we can easily check that

\begin{subequations}\label{29}\begin{eqnarray}
\frac{d{\widetilde E}}{dt} &=& -\frac{1}{5 c^5} {\dddot Q}_{ij}
{\dddot Q}_{ij} + {\cal O}\left( \frac{1}{c^7} \right)\;, \\
\frac{d{\widetilde J}^i}{dt} &=& - \frac{2}{5 c^5}\varepsilon_{ijk}
{\ddot Q}_{jl} {\dddot Q}_{kl} + {\cal O}\left( \frac{1}{c^7}
\right)\;,
\end{eqnarray}\end{subequations}
where the Newtonian trace-free quadrupole moment is $Q_{ij}=\mu (x^i
x^j-\frac{1}{3}\delta^{ij} r^2)$.

\section{Lagrangian and Hamiltonian in ADM-coordinates}

In Ref. \cite{ABF01} (see also Ref. \cite{DJS01}) we determined the
``contact'' transformation between the particles' variables in
harmonic coordinates and those in ADM (or rather
ADM-type\footnote{Strictly speaking, the ADM coordinates we are
considering differ from the actual ADM coordinates at the 3PN order by
a shift of phase-space coordinates that is given in
Ref. \cite{DJS00}.}) coordinates. By contact transformation we mean
the relation between the particles' trajectories in harmonic
coordinates, $y_A^i(t)$, and the corresponding ones in ADM
coordinates, say $Y_A^i(t)$. We recall that in the contact
transformation, i.e.

\begin{equation}\label{29''}
\delta y_A^i(t)=Y_A^i(t)-y_A^i(t)\;,
\end{equation}
the time coordinate $t$ is to be viewed as a ``dummy''
variable\footnote{The contact transformation is not a coordinate
transformation between the spatial vectors in both coordinates, but
takes also into account the fact that the time coordinate changes as
well~: i.e. $\delta y_A^i=\xi^i(y_A)-\xi^0(y_A)v_A^i/c$, where
$\xi^\mu(y_A)$ denotes the four-dimensional change between the
harmonic and ADM coordinates, when evaluated at the position
$y_A=(t,y_A^i)$.}. There is a unique transformation (\ref{29''}) such
that the 3PN harmonic-coordinates Lagrangian of de Andrade, Blanchet
and Faye \cite{ABF01} (in a general frame) is changed into another
Lagrangian whose Legendre transform coincides with the 3PN
ADM-coordinates Hamiltonian derived by Damour, Jaranowski and
Sch\"afer \cite{DJS00}. The explicit expression of this general-frame
contact transformation can be found in Section 4.2 of
Ref. \cite{ABF01}.

Now we are in a position to obtain the relation between the relative
separation vector $x^i\equiv y_1^i-y_2^i$ in harmonic coordinates and
the one $X^i\equiv Y_1^i-Y_2^i$ in ADM coordinates (do not confuse the
relative distances $x^i$ and $X^i$ between the two particles with the
spatial position vector of some field event in these
coordinates). Namely,

\begin{equation}\label{30'}
\delta x^i\equiv X^i-x^i=\delta y_1^i-\delta y_2^i\;,
\end{equation}
where $\delta y_1^i$ and $\delta y_2^i$ are given by Eqs. (4.8)-(4.10)
in Ref. \cite{ABF01}. [We shall always view such equalities as
(\ref{30'}) as functional equalities, i.e. valid for any dummy time
variable $t$.]  One replaces in Eq. (\ref{30'}) all the variables by
their center-of-mass counterparts following
Eqs. (\ref{21})-(\ref{22}). Actually, since the contact transformation
is already of relative order 2PN, the calculation is quite immediate
and requires only the equations (\ref{21})-(\ref{22}) to 1PN order. As
a result the spatial separation vectors $X^i$ and $x^i$ in both
coordinates (each one living within the spatial slice of constant
time appropriate to each of the coordinate systems) are related to
each other by

\begin{eqnarray}\label{31'}
\frac{\delta x^i}{m} &=&
\frac{1}{c^4}\bigg\{\left[\frac{\dot{r}^2\,\nu}{8} -
\frac{5\,\nu\,v^2}{8}+\frac{m}{r}\left( -\frac{1}{4}- 3\,\nu \right)
\right] n^i +\frac{9\,\dot{r}\,\nu}{4} v^i\bigg\} \nonumber\\ &+&
\frac{1}{c^6}\bigg\{\left[-\frac{\dot{r}^4\,\nu}{16} +
\frac{5\,\dot{r}^4\,\nu^2}{16} + \frac{5\,\dot{r}^2\,\nu\,v^2}{16} -
\frac{15\,\dot{r}^2\,\nu^2\,v^2}{16} - \frac{\nu\,v^4}{2} +
\frac{11\,\nu^2\,v^4}{8} \right.\nonumber\\&&\qquad~ +
\left.\frac{m}{r}\left( \frac{161\,\dot{r}^2\,\nu}{48} -
\frac{5\,\dot{r}^2\,\nu^2}{2} - \frac{451\,\nu\,v^2}{48} -
\frac{3\,\nu^2\,v^2}{8} \right) \right.\nonumber\\&& \qquad~ +
\left. \frac{m^2}{r^2} \left( \frac{2773\,\nu}{280} +
\frac{21\,\nu\,\pi^2}{32} - \frac{22\,\nu\,}{3}\ln
\Big(\frac{r}{r'_0}\Big) \right)\right] n^i
\nonumber\nonumber\\&&\quad +\left[-\frac{5\,\dot{r}^3\,\nu}{12} +
\frac{29\,\dot{r}^3\,\nu^2}{24} + \frac{17\,\dot{r}\,\nu\,v^2}{8} -
\frac{21\,\dot{r}\,\nu^2\,v^2}{4} \right.\nonumber\\&&\qquad~ +
\left. \frac{m}{r}\left( \frac{43\,\dot{r}\,\nu}{3} +
5\,\dot{r}\,\nu^2 \right)\right] v^i\bigg\}\;.
\end{eqnarray}
Below we shall deduce from this formula the radius of the circular
orbit in ADM coordinates, say $R_0$, versus the one in
harmonic-coordinates, i.e. $r_0$; see Eq. (\ref{54}).

Now we look for the center-of-mass Lagrangian in ADM
coordinates. Since in ADM coordinates the Lagrangian is ``ordinary''
(no accelerations) the contact transformation must be such that it
removes the acceleration terms present in harmonic coordinates ---
more precisely, it must make them in the form of a total time
derivative which is irrelevant to the dynamics. Following an
investigation similar to the one in Section 3.2 of Ref. \cite{ABF01}
[see notably Eq. (3.18) there] we know that ${\cal L}^{\rm ADM}$
differs from ${\cal L}$ by two terms~: (1) the functional variation of
${\cal L}$ induced at the linearized order by the contact
transformation of the relative path as given by
Eqs. (\ref{30'})-(\ref{31'}); (2) the total time derivative of a
function ${\cal F}$ of the relative position and velocity. We can
limit our consideration to the linearized order because $\delta x^i$
is at least of order 2PN, so the non-linear terms do not contribute
before the 4PN order and are negligible here. Hence we necessarily
have the following {\it functional} equality (by which we mean the
equality between functions of the same {\it dummy} variables ${\bf
x}$, ${\bf v}$, ${\bf a}$)~:

\begin{equation}\label{32}
{\cal L}^{\rm ADM}[{\bf x},{\bf v}] = {\cal L}[{\bf x},{\bf v}, {\bf
a}]-\frac{\delta {\cal L}}{\delta x^i}\delta x^i+\frac{d {\cal F}}{dt}\;,
\end{equation}
in which $\delta x^i$ is explicitly given by Eq. (\ref{31'}), and
where the function ${\cal F}$ is for the moment unknown\footnote{Note
that the minus sign in front of the second term in Eq. (\ref{32})
differs from the one in Eq. (3.18) of Ref. \cite{ABF01}. The reason is
because we have corrected a sign inconsistency in Ref. \cite{ABF01}~:
namely the equation (3.12) there, together with the adopted definition
$\delta y_A^i={y'_A}^i-y_A^i$, is inconsistent with (3.13) and (3.18);
but this does not change any of the results of Ref. \cite{ABF01}.}. We
insist that in the present calculation the contact transformation
$\delta x^i$ is known so that the only freedom left is the choice of
${\cal F}$. This contrasts with our earlier study in Ref. \cite{ABF01}
where both the contact transformation of the individual paths, $\delta
y_1^i$ and $\delta y_2^i$, and some arbitrary function, say $F$, had
to be varied and determined. The reason of course is that once $\delta
y_1^i$ and $\delta y_2^i$ are known from Ref. \cite{ABF01} we have no
choice for $\delta x^i$ which must be equal to the center-of-mass
reduction of the difference $\delta y_1^i-\delta y_2^i$ [see
Eq. (\ref{30'})]. Thus, despite the smaller freedom that we presently
have in the adjustment of parameters, the calculation {\it must} work
with that $\delta x^i$ and not with another one.

The function ${\cal F}={\cal F}[{\bf x},{\bf v}]$ is not difficult to
determine in order to match perfectly the ADM Hamiltonian. Notice that
after adding the total time derivative of that function, not only has
one been able to remove all the accelerations, but also one has gauged
away all the logarithms which were present in the harmonic-coordinates
Lagrangian. We get

\begin{eqnarray}\label{33}
\frac{{\cal F}}{m\,\dot{r}} &=&\frac{1}{c^4}\bigg[ - \frac{\nu\,v^2}{4}
+ \frac{m}{r}\,\left( \frac{1}{4} + 3\,\nu \right)\bigg]\nonumber\\
&+& \frac{1}{c^6}\bigg[-\frac{\nu\,\dot{r}^2\,v^2}{16} +
\frac{19\,\nu^2\,\dot{r}^2\,v^2}{48} - \frac{\nu\,v^4}{16} +
\frac{19\,\nu^2\,v^4}{16} \nonumber\\&&\qquad + \frac{m}{r}\,\left(
-\frac{43\,\nu\,\dot{r}^2}{144} - \frac{97\,\nu^2\,\dot{r}^2}{36} +
\frac{v^2}{8} - \frac{217\,\nu\,v^2}{48} - \frac{665\,\nu^2\,v^2}{24}
\right) \nonumber\\&&\qquad + \frac{m^2}{r^2}\,\left( \frac{3}{4} -
\frac{113\,\nu}{280} + 6\,\nu^2 - \frac{21\,\nu\,\pi^2}{32} +
\frac{22\,\nu}{3}\ln \Big(\frac{r}{r'_0}\Big) \right)\bigg]\;.
\end{eqnarray}
Next, Eq. (\ref{32}) together with the explicit expressions
(\ref{31'}) and (\ref{33}) give the ADM center-of-mass
Lagrangian. Once the calculation is done we have to express it using
the names appropriate to the ADM variables~: $X^i=x^i+\delta x^i$,
which means the separation distance $R$, the relative square velocity
$V^2$, and the radial velocity $\dot{R}={\bf N}\cdot{\bf V}$. The formula
is

\begin{eqnarray}\label{34}
\frac{{\cal L}^{\rm ADM}}{\mu} &=&\frac{m}{R}+ \frac{V^2}{2}
\nonumber\\ && +\frac{1}{c^2}\bigg\{\frac{V^4}{8} -
\frac{3\,\nu\,V^4}{8} + \frac{m}{R}\,\left( \frac{\nu\,\dot{R}^2}{2} +
\frac{3\,V^2}{2} + \frac{\nu\,V^2}{2} \right)-\frac{m^2}{2R^2}
\bigg\}\nonumber\\ && +\frac{1}{c^4}\bigg\{\frac{V^6}{16} -
\frac{7\,\nu\,V^6}{16} + \frac{13\,\nu^2\,V^6}{16}
\nonumber\\&&\qquad~+ \frac{m}{R}\,\left(
\frac{3\,\nu^2\,\dot{R}^4}{8} + \frac{\nu\,\dot{R}^2\,V^2}{2} -
\frac{5\,\nu^2\,\dot{R}^2\,V^2}{4} + \frac{7\,V^4}{8} -
\frac{3\,\nu\,V^4}{2} - \frac{9\,\nu^2\,V^4}{8}
\right)\nonumber\\&&\qquad~ + \frac{m^2}{R^2}\,\left(
\frac{3\,\nu\,\dot{R}^2}{2} + \frac{3\,\nu^2\,\dot{R}^2}{2} + 2\,V^2 -
\nu\,V^2 + \frac{\nu^2\,V^2}{2} \right) \nonumber\\&&\qquad~
+\frac{m^3}{R^3}\,\left( \frac{1}{4} + \frac{3\,\nu}{4}
\right)\bigg\}\nonumber\\ && +\frac{1}{c^6}\bigg\{ \frac{5\,V^8}{128}
- \frac{59\,\nu\,V^8}{128} + \frac{119\,\nu^2\,V^8}{64} -
\frac{323\,\nu^3\,V^8}{128}\nonumber\\&&\qquad~+\frac{m}{R}\,\left(
\frac{5\,\nu^3\,\dot{R}^6}{16} + \frac{9\,\nu^2\,\dot{R}^4\,V^2}{16} -
\frac{33\,\nu^3\,\dot{R}^4\,V^2}{16} + \frac{\nu\,\dot{R}^2\,V^4}{2} -
3\,\nu^2\,\dot{R}^2\,V^4 \right.\nonumber\\&&\qquad\qquad\quad~\left.
+ \frac{75\,\nu^3\,\dot{R}^2\,V^4}{16} + \frac{11\,V^6}{16} -
\frac{7\,\nu\,V^6}{2} + \frac{59\,\nu^2\,V^6}{16} +
\frac{65\,\nu^3\,V^6}{16} \right)\nonumber\\&&\qquad~
+\frac{m^2}{R^2}\,\left( - \frac{5\,\nu\,\dot{R}^4}{12} +
\frac{17\,\nu^2\,\dot{R}^4}{12} + 2\,\nu^3\,\dot{R}^4 +
\frac{39\,\nu\,\dot{R}^2\,V^2}{16} -
\frac{29\,\nu^2\,\dot{R}^2\,V^2}{8}
\right.\nonumber\\&&\qquad\qquad\quad~\left. -
\frac{27\,\nu^3\,\dot{R}^2\,V^2}{4} + \frac{47\,V^4}{16} -
\frac{15\,\nu\,V^4}{4} - \frac{25\,\nu^2\,V^4}{16} -
\frac{9\,\nu^3\,V^4}{4} \right) \nonumber\\&&\qquad~
+\frac{m^3}{R^3}\,\left( \frac{77\,\nu\,\dot{R}^2}{16} +
\frac{5\,\nu^2\,\dot{R}^2}{4} + \frac{7\,\nu^3\,\dot{R}^2}{2} +
\frac{3\,\nu\,\dot{R}^2\,\pi^2}{64} + \frac{13\,V^2}{8}
\right.\nonumber\\&&\qquad\qquad\quad~\left. -
\frac{409\,\nu\,V^2}{48} - \frac{5\,\nu^2\,V^2}{8} +
\frac{\nu^3\,V^2}{2} - \frac{\nu\,{\pi }^2\,V^2}{64} \right)
\nonumber\\&&\qquad~ +\frac{m^4}{R^4}\,\left( -\frac{1}{8} -
\frac{1881\,\nu}{280} + \frac{11\,\lambda\,\nu}{3} +
\frac{21\,\nu\,\pi^2}{32} \right) \bigg\}\;.
\end{eqnarray}
This is an ordinary Lagrangian and we apply the ordinary Legendre
transform to obtain the Hamiltonian, which is a function of the
conjugate momentum

\begin{subequations}\label{34'}\begin{eqnarray}
P^i &=& \frac{\partial {\cal L}^{\rm ADM}}{\partial
V^i}\;,\label{34a'}\\ P^2 &\equiv& {\bf P}^2\quad\hbox{and}\quad P_R
\equiv {\bf N}\cdot{\bf P}\;.
\end{eqnarray}\end{subequations}
We find perfect agreement with the center-of-mass Hamiltonian derived
in Refs. \cite{JaraS98,JaraS99,DJS00}~:

\begin{eqnarray}\label{35}
\frac{{\cal H}^{\rm ADM}}{\mu} &=& \frac{P^2}{2}
-\frac{m}{R}\nonumber\\ && +\frac{1}{c^2}\bigg\{- \frac{P^4}{8} +
\frac{3\,\nu\,P^4}{8} + \frac{m}{R}\left( - \frac{{P_R}^2\,\nu}{2} -
\frac{3\,P^2}{2} - \frac{\nu\,P^2}{2} \right)+\frac{m^2}{2R^2}
\bigg\}\nonumber\\ && +\frac{1}{c^4}\bigg\{\frac{P^6}{16} -
\frac{5\,\nu\,P^6}{16} + \frac{5\,\nu^2\,P^6}{16}
\nonumber\\&&\qquad~+ \frac{m}{R}\left( - \frac{3\,{P_R}^4\,\nu^2}{8}
- \frac{{P_R}^2\,P^2\,\nu^2}{4} + \frac{5\,P^4}{8} -
\frac{5\,\nu\,P^4}{2} - \frac{3\,\nu^2\,P^4}{8}
\right)\nonumber\\&&\qquad~ + \frac{m^2}{R^2}\,\left(
\frac{3\,{P_R}^2\,\nu}{2} + \frac{5\,P^2}{2} + 4\,\nu\,P^2 \right)
\nonumber\\&&\qquad~+\frac{m^3}{R^3}\left( -\frac{1}{4} -
\frac{3\,\nu}{4} \right) \bigg\}\nonumber\\ &&
+\frac{1}{c^6}\bigg\{-\frac{5\,P^8}{128} + \frac{35\,\nu\,P^8}{128} -
\frac{35\,\nu^2\,P^8}{64} + \frac{35\,\nu^3\,P^8}{128}
\nonumber\\&&\qquad~ + \frac{m}{R}\left( -\frac{5\,{P_R}^6\,\nu^3}{16}
+ \frac{3\,{P_R}^4\,P^2\,\nu^2}{16} -
\frac{3\,{P_R}^4\,P^2\,\nu^3}{16} + \frac{{P_R}^2\,P^4\,\nu^2}{8}
\right.\nonumber\\&&\qquad\qquad\quad~ \left. -
\frac{3\,{P_R}^2\,P^4\,\nu^3}{16}-\frac{7\,P^6}{16} +
\frac{21\,\nu\,P^6}{8} - \frac{53\,\nu^2\,P^6}{16} -
\frac{5\,\nu^3\,P^6}{16} \right) \nonumber\\&&\qquad~ +
\frac{m^2}{R^2}\,\left( \frac{5\,{P_R}^4\,\nu}{12} +
\frac{43\,{P_R}^4\,\nu^2}{12} + \frac{17\,{P_R}^2\,P^2\,\nu}{16}
\right.\nonumber\\&&\qquad\qquad\quad~ \left.+
\frac{15\,{P_R}^2\,P^2\,\nu^2}{8} - \frac{27\,P^4}{16} +
\frac{17\,\nu\,P^4}{2} + \frac{109\,\nu^2\,P^4}{16} \right)
\nonumber\\&&\qquad~ + \frac{m^3}{R^3}\,\left(
-\frac{85\,{P_R}^2\,\nu}{16} - \frac{7\,{P_R}^2\,\nu^2}{4} -
\frac{25\,P^2}{8} - \frac{335\,\nu\,P^2}{48}
\right.\nonumber\\&&\qquad\qquad\quad~ \left.-
\frac{23\,\nu^2\,P^2}{8} - \frac{3\,{P_R}^2\,\nu\,\pi^2}{64} +
\frac{\nu\,P^2\,\pi^2}{64} \right)\nonumber\\&&\qquad~ +
\frac{m^4}{R^4}\, \left( \frac{1}{8} + \frac{1881\,\nu}{280} -
\frac{11\,\lambda\,\nu}{3} - \frac{21\,\nu\,\pi^2}{32} \right) \bigg\}\;.
\end{eqnarray}
Recall that $\lambda$ is related to the so-called ``static''
regularization-ambiguity constant $\omega_{\rm s}$ of
Refs. \cite{JaraS98,JaraS99} by $\lambda=-\frac{3}{11}\omega_{\rm
s}-\frac{1987}{3080}$. We have $\omega_{\rm s}=0$ according to
Ref. \cite{DJSdim}. On the other hand, the ``kinetic'' ambiguity
constant $\omega_{\rm k}$ of Refs. \cite{JaraS98,JaraS99} has been
fixed to the value $\omega_{\rm k}=\frac{41}{24}$ by the explicit
Lorentz invariance of the equations of motion in harmonic coordinates
\cite{BF00,BFeom}, and, equivalently, by the requirement of existence
of generators for the Poincar\'e algebra in the ADM-Hamiltonian
formalism \cite{DJS00}.

Finally let us present, for completeness, the formulas for the
center-of-mass positions which are analogous to
Eqs. (\ref{21})-(\ref{22}) but in ADM coordinates. We have

\begin{subequations}\label{35'}\begin{eqnarray}
Y_1^i&=&\Big[X_2+\nu (X_1-X_2) \hat{\cal P}\Big] X^i +\nu
(X_1-X_2)\hat{\cal Q}\,V^i+ {\cal O}\left( \frac{1}{c^7}
\right)\;,\label{35a'}\\ Y_2^i&=&\Big[-X_1+\nu (X_1-X_2) \hat{\cal
P}\Big] X^i +\nu (X_1-X_2)\hat{\cal Q}\,V^i+ {\cal O}\left(
\frac{1}{c^7} \right)\;.\label{35b'}
\end{eqnarray}\end{subequations}
where the post-Newtonian coefficients $\hat{\cal P}$ and $\hat{\cal
Q}$ are given by

\begin{subequations}\label{35''}\begin{eqnarray}
\hat{\cal P}&=&\frac{1}{c^2}\bigg[\frac{V^2}{2}
-\frac{m}{2R}\bigg]\nonumber\\ &+& \frac{1}{c^4}\bigg[
\frac{3\,V^4}{8} - \frac{3\,\nu\,V^4}{2} \nonumber\\ && \qquad+
\frac{m}{R}\,\left( \frac{3\,\nu\,\dot{R}^2}{4} + \frac{7\,V^2}{4} +
\frac{3\,\nu\,V^2}{2} \right)\nonumber\\ &&
\qquad+\frac{m^2}{R^2}\left(\frac{1}{4} - \frac{\nu}{2} \right)
\bigg]\nonumber\\ &+& \frac{1}{c^6}\bigg[\frac{5\,V^6}{16} -
\frac{11\,\nu\,V^6}{4} + 6\,\nu^2\,V^6 \nonumber\\ && \qquad
+\frac{m}{R}\left( \frac{21\,\dot{R}^4\,\nu^2}{16} +
\frac{7\,\dot{R}^2\,\nu\,V^2}{4} -
\frac{11\,\dot{R}^2\,\nu^2\,V^2}{2}\right.\nonumber\\ &&
\qquad\qquad\quad +\left.  \frac{45\,V^4}{16} -
\frac{109\,\nu\,V^4}{16} - \frac{15\,\nu^2\,V^4}{2} \right)
\nonumber\\ && \qquad +\frac{m^2}{R^2}\left(
\frac{9\,\dot{R}^2\,\nu}{4} + 4\,\dot{R}^2\,\nu^2 + \frac{23\,V^2}{8}
+ \frac{29\,\nu\,V^2}{16} + 3\,\nu^2\,V^2 \right) \nonumber\\ &&
\qquad + \frac{m^3}{R^3}\left( -\frac{1}{8} + \frac{\nu}{8} -
\frac{\nu^2}{2} \right) \bigg]\;,\\ \hat{\cal
Q}&=&\frac{1}{c^5}\bigg[\frac{4\,m\,V^2}{5} -\frac{8\,m^2}{5\,R} \bigg]
+ \frac{1}{c^6}\bigg[ \frac{m^2\,\nu\,\dot{R}}{4\,R}\bigg]\;.
\label{35b''}\end{eqnarray}\end{subequations}
At the 2PN order the result is identical with the one given by Wex in
his Appendix A \cite{W95}\footnote{Since the contact transformation we
consider relates together the {\it conservative} parts of the dynamics
in harmonic and ADM-type coordinates (and affects only the 2PN and 3PN
orders), the radiation-reaction damping term at the 2.5PN order in
Eq. (\ref{35b''}) is the same as in harmonic coordinates. This is
merely a definition of a particular ADM-type dynamics ({\it a priori}
different from the one in ADM coordinates {\it stricto-sensu}), in
which the ``odd'' terms, associated with radiation reaction, are the
same as in harmonic coordinates.}. By differentiating Eqs. (\ref{35'})
with respect to time we obtain the center-of-mass velocities $V_1^i$
and $V_2^i$ in terms of the relative position $X^i$ and velocity
$V^i$. We have checked that by replacing into the obtained relations
the velocities $V_1^i$ and $V_2^i$ by their expressions depending on
the conjugate momenta $P_1^i$ and $P_2^i$ as deduced from the
variation of the general-frame Lagrangian, and by expressing the
velocity $V^i$ in terms of $P^i$ following the variation of the
center-of-mass Lagrangian [Eq. (\ref{34a'})], with both replacements
being made with the full 3PN accuracy, one ends up with the simple
equations

\begin{equation}
P_1^i = P^i = -P_2^i\;,
\end{equation}
which are indeed the ones appropriate to a linear momentum that is
conserved.

\section{On the dynamical stability of circular orbits}

As an application of the previous formalism let us investigate the
problem of the stability, against dynamical perturbations, of circular
orbits at the 3PN order. We propose to use two different methods, one
based on a perturbation at the level of the equations of motion
(\ref{24})-(\ref{25}) in harmonic coordinates, the other one
consisting of perturbing the Hamiltonian equations in ADM coordinates
for the Hamiltonian (\ref{35}). We shall find a criterion for the
stability of orbits and shall present it in an invariant way (the same
in different coordinate systems). We shall check that our two methods
agree on the result.

We deal first with the perturbation of the equations of motion,
following Kidder, Will and Wiseman \cite{KWW93} (see their Section
III.A). We introduce polar coordinates $(r,\varphi)$ in the orbital
plane and pose $u\equiv {\dot r}$ and $\omega\equiv {\dot \varphi}$
(beware that in this paper $u= \dot{r}$, and {\it not} another
standard notation in central force problems, $u= 1/r$).  Then
Eq. (\ref{24}) yields the system of equations

\begin{subequations}\label{36}\begin{eqnarray}
{\dot r} &=& u\;,\label{36a}\\ {\dot u} &=& -\frac{m}{r^2}\Big[1+{\cal
A}+{\cal B}u\Big]+r\omega^2\;,\label{36b}\\ {\dot \omega} &=&
-\omega\left[\frac{m}{r^2}{\cal B}+\frac{2 u}{r}\right]\;,\label{36c}
\end{eqnarray}\end{subequations}
where ${\cal A}$ and ${\cal B}$ are given by Eqs. (\ref{25}) as
functions of $r$, $u$ and $\omega$ (through $v^2=u^2+r^2\omega^2$). In
the case of an orbit which is circular apart from the adiabatic
inspiral at the 2.5PN order (we neglect the 2.5PN radiation-reaction
effect), we have $\dot r=\dot u=\dot \omega=0$ hence
$u=0$. Eq. (\ref{36b}) gives thereby the angular velocity $\omega_0$
of the circular orbit as

\begin{equation}\label{37}
\omega_0^2 = \frac{m}{r_0^3}\big(1+{\cal A}_0\big)\;.
\end{equation}
Solving iteratively this relation at the 3PN order using the equations
of motion (\ref{24})-(\ref{25}) we obtain $\omega_0$ as a function of
the circular-orbit radius $r_0$ in harmonic coordinates (the result
agrees with the one of Refs. \cite{BF00,BFeom})~:

\begin{eqnarray}\label{42}
\omega_0^2 =
\frac{m}{r_0^3}\Bigg\{1&+&\frac{m}{r_0\,c^2}\Big(-3+\nu\Big)
+\frac{m^2}{r_0^2\,
c^4}\left(6+\frac{41}{4}\nu+\nu^2\right)\nonumber\\
&+&\frac{m^3}{r_0^3\,c^6}\left(-10
+\left[-\frac{67759}{840}+\frac{41}{64}\pi^2+22\ln
\Big(\frac{r_0}{r'_0}\Big)+\frac{44}{3}\lambda
\right]\nu+\frac{19}{2}\nu^2+\nu^3\right)\nonumber\\ &+&{\cal
O}\left(\frac{1}{c^8}\right)\Bigg\}\;.
\end{eqnarray}
[Please do not confuse the circular-orbit radius $r_0$ with the
constant $r'_0$ entering the logarithm at the 3PN order and which is
defined by Eq. (\ref{26}).]

Now we investigate the equations of linear perturbations around the
circular orbit defined by the constants $r_0$, $u_0=0$ [actually, if
we were to include the radiation-reaction damping, $u_0={\cal
O}(c^{-5})$] and $\omega_0$. We pose

\begin{subequations}\label{38}\begin{eqnarray}
r &=& r_0 + \delta r\;,\\
u &=& \delta u\;,\\
\omega &=& \omega_0 + \delta \omega\;,
\end{eqnarray}\end{subequations}
where $\delta r$, $\delta u$ and $\delta \omega$ denote some
perturbations of the circular orbit. Then a system of linear equations
follows~:

\begin{subequations}\label{39}\begin{eqnarray}
\dot{\delta r} &=& \delta u\;,\\
\dot{\delta u} &=& \alpha_0\, \delta r + \beta_0\, \delta \omega\;,\\
\dot{\delta \omega} &=& \gamma_0\, \delta u\;,
\end{eqnarray}\end{subequations}
where the coefficients, which solely depend on the unperturbed
circular orbit, read \cite{KWW93}

\begin{subequations}\label{40}\begin{eqnarray}
\alpha_0 &=& 3 \omega_0^2 - \frac{m}{r_0^2}\left(\frac{\partial {\cal
A}}{\partial r}\right)_0\;,\\ \beta_0 &=& 2 r_0 \omega_0 -
\frac{m}{r_0^2}\left(\frac{\partial {\cal A}}{\partial
\omega}\right)_0\;,\\ \gamma_0 &=& -\omega_0 \left[\frac{2}{r_0} +
\frac{m}{r_0^2}\left(\frac{\partial {\cal B}}{\partial u}\right)_0\right]\;.
\end{eqnarray}\end{subequations}
In obtaining Eqs. (\ref{40}) we use the fact that ${\cal A}$ is a
function of the square $u^2$ through $v^2=u^2+r^2\omega^2$, so that
$\partial {\cal A}/\partial u$ is proportional to $u$ and thus
vanishes in the unperturbed configuration (because $u=\delta u$). On
the other hand, since the radiation reaction is neglected, ${\cal B}$
also is proportional to $u$ [see Eq. (\ref{25b})], so only $\partial
{\cal B}/\partial u$ can contribute at the zeroth perturbative
order. Now by examining the fate of perturbations that are
proportional to some $e^{i\sigma t}$, we arrive at the condition for
the frequency $\sigma$ of the perturbation to be real, and hence for
stable circular orbits to exist, as being \cite{KWW93}

\begin{equation}\label{41}
\hat{C}_0 \equiv -\alpha_0 - \beta_0\, \gamma_0 ~> 0\;.
\end{equation}
Substituting into this ${\cal A}$ and ${\cal B}$ at the 3PN order we
then arrive at the orbital-stability criterion

\begin{eqnarray}\label{43}
\hat{C}_0 =
\frac{m}{r_0^3}\Bigg\{1&+&\frac{m}{r_0\,c^2}\Big(-9+\nu\Big)
+\frac{m^2}{r_0^2\,c^4}\left(30
+\frac{65}{4}\nu+\nu^2\right)\nonumber\\ &+&\frac{m^3}{r_0^3\,c^6}
\left(-70+\left[-\frac{45823}{840}-\frac{451}{64}\pi^2+22\ln
\Big(\frac{r_0}{r'_0}\Big)-\frac{88}{3}\lambda
\right]\nu+\frac{19}{2}\nu^2+\nu^3\right)\nonumber\\ &+&{\cal
O}\left(\frac{1}{c^8}\right)\Bigg\}\;,
\end{eqnarray}
where we recall that $r_0$ is the radius of the orbit in harmonic
coordinates.

Our second method is to use the Hamiltonian equations based on the 3PN
Hamiltonian in ADM coordinates given by Eq. (\ref{35}). We introduce
the polar coordinates $(R,\Psi)$ in the orbital plane --- we assume
that the orbital plane is equatorial, given by $\Theta=\frac{\pi}{2}$
in the spherical coordinate system $(R,\Theta,\Psi)$ --- and make the
substitution

\begin{equation}\label{44}
P^2={P_R}^2+\frac{P_\Psi^2}{R^2}\;,
\end{equation}
into the Hamiltonian. This yields a ``reduced'' Hamiltonian that is a
function of $R$, $P_R$ and $P_\Psi$~: ${\cal H}={\cal
H}\big[R,P_R,P_\Psi\big]$, and describes the motion in polar
coordinates in the orbital plane (henceforth we denote ${\cal H}={\cal
H}^{\rm ADM}/\mu$). The Hamiltonian equations then read

\begin{subequations}\label{45}\begin{eqnarray}
\frac{dR}{dt} &=& \frac{\partial {\cal H}}{\partial P_R}\;,\\
\frac{d\Psi}{dt} &=& \frac{\partial {\cal H}}{\partial P_\Psi}\;,\\
\frac{dP_R}{dt} &=& -\frac{\partial {\cal H}}{\partial R}\;,\\
\frac{dP_\Psi}{dt} &=& 0\;.
\end{eqnarray}\end{subequations}
Evidently the constant $P_\Psi$ is nothing but the conserved
angular-momentum integral. For circular orbits we have $R=R_0$ (a
constant) and $P_R=0$, so

\begin{equation}\label{46}
\frac{\partial {\cal H}}{\partial R}\big[R_0,0,P_\Psi^0\big] = 0\;,
\end{equation}
which gives the angular momentum $P_\Psi^0$ of the circular orbit as a
function of $R_0$, and

\begin{equation}\label{47}
\omega_0 \equiv \left(\frac{d\Psi}{dt}\right)_0 = \frac{\partial {\cal
H}}{\partial P_\Psi}\big[R_0,0,P_\Psi^0\big]\;,
\end{equation}
which yields the angular frequency of the circular orbit $\omega_0$
--- the same as in Eq. (\ref{42}) --- in terms of $R_0$ \footnote{The
last equation,
$$
\frac{\partial {\cal H}}{\partial P_R}\big[R_0,0,P_\Psi^0\big] = 0\;,
$$
which is equivalent to $R={\rm const}=R_0$, is automatically verified
because ${\cal H}$ is a quadratic function of $P_R$ and hence
$\partial {\cal H}/\partial P_R$ is zero for circular orbits.}~:

\begin{eqnarray}
\omega_0^2 =
\frac{m}{R_0^3}\Bigg\{1&+&\frac{m}{R_0\,c^2}\Big(-3+\nu\Big)
+\frac{m^2}{R_0^2\,
c^4}\left(\frac{21}{4}-\frac{5}{8}\nu+\nu^2\right)\nonumber\\
&+&\frac{m^3}{R_0^3\,c^6}\left(-7
+\left[-\frac{54629}{1680}+\frac{167}{64}\pi^2
+\frac{44}{3}\lambda
\right]\nu-\frac{31}{8}\nu^2+\nu^3\right)\nonumber\\ &+&{\cal
O}\left(\frac{1}{c^8}\right)\Bigg\}\;.
\end{eqnarray}

We consider now a perturbation of the circular orbit defined by

\begin{subequations}\label{49}\begin{eqnarray}
P_R &=& \delta P_R\;,\\ 
P_\Psi &=& P_\Psi^0 + \delta P_\Psi\;,\\
R &=& R_0 + \delta R\;,\\ 
\omega &=& \omega_0 + \delta \omega\;.
\end{eqnarray}\end{subequations}
It is easy to verify that the Hamiltonian equations (\ref{45}), when
worked out at the linearized order, read as

\begin{subequations}\label{50}\begin{eqnarray}
\dot{\delta P_R} &=& -\pi_0\, \delta R - \rho_0\, \delta P_\Psi\;,\\
\dot{\delta P_\Psi} &=& 0\;,\\
\dot{\delta R} &=& \sigma_0\, \delta P_R\;,\\
\delta \omega &=& \rho_0\, \delta R + \tau_0\, \delta P_\Psi\;,
\end{eqnarray}\end{subequations}
where the coefficients, which depend on the unperturbed orbit, are
given by

\begin{subequations}\label{51}\begin{eqnarray}
\pi_0&=&\frac{\partial^2 {\cal H}}{\partial R^2}\big[R_0,0,P_\Psi^0\big]\;,\\
\rho_0&=&\frac{\partial^2 {\cal H}}{\partial R\, \partial
P_\Psi}\big[R_0,0,P_\Psi^0\big]\;,\\ \sigma_0&=&\frac{\partial^2 {\cal
H}}{\partial {P_R}^2}\big[R_0,0,P_\Psi^0\big]\;,\\ \tau_0&=&\frac{\partial^2
{\cal H}}{\partial {P_\Psi}^2}\big[R_0,0,P_\Psi^0\big]\;.
\end{eqnarray}\end{subequations}
By looking to solutions proportional to some $e^{i\sigma t}$ one
obtains some real frequencies, and therefore one finds stable circular
orbits, if and only if

\begin{equation}\label{52}
\hat{C}_0 \equiv  \pi_0\, \sigma_0 ~> 0\;.
\end{equation}
Using the Hamiltonian (\ref{35}) we readily obtain

\begin{eqnarray}\label{53}
\hat{C}_0 = \frac{m}{R_0^3}\Bigg\{1&+&\frac{m}{R_0\,c^2}(-9+\nu)
+\frac{m^2}{R_0^2\,c^4}\left(\frac{117}{4}
+\frac{43}{8}\nu+\nu^2\right)\nonumber\\
&+&\frac{m^3}{R_0^3\,c^6}\left(-61+\left[\frac{135403}{1680}
-\frac{325}{64}\pi^2-\frac{88}{3}\lambda
\right]\nu-\frac{31}{8}\nu^2+\nu^3\right)\nonumber\\ &+&{\cal
O}\left(\frac{1}{c^8}\right)\Bigg\}\;.
\end{eqnarray}
This result does not look the same as our previous result (\ref{43}),
but this is simply due to the fact that it depends on the ADM radial
separation $R_0$ instead of the harmonic one $r_0$. Fortunately we
have derived in Section V all the material needed to connect $R_0$ to
$r_0$ with the 3PN accuracy. Indeed, with Eqs. (\ref{30'})-(\ref{31'})
we have the relation valid for general orbits between the separation
vectors in both coordinate systems. Specializing that relation to
circular orbits we readily find

\begin{eqnarray}\label{54}
R_0 = r_0\Bigg\{1&+&\frac{m^2}{r_0^2\,
c^4}\left(-\frac{1}{4}-\frac{29}{8}\nu\right)
+\frac{m^3}{r_0^3\, c^6}\left(\left[\frac{3163}{1680}
+\frac{21}{32}\pi^2
-\frac{22}{3}\ln\Big(\frac{r_0}{r'_0}\Big)\right]\nu 
+\frac{3}{8}\nu^2\right)\nonumber\\ &+&{\cal
O}\left(\frac{1}{c^8}\right)\Bigg\}\;.
\end{eqnarray}
The difference between $R_0$ and $r_0$ is made out of 2PN and 3PN
terms only. Inserting Eq. (\ref{54}) into Eq. (\ref{53}) and
re-expanding to 3PN order we find that indeed our basic
stability-criterion function $\hat{C}_0$ comes out the same with our
two methods.

Finally let us give to the function $\hat{C}_0$ an invariant meaning by
expressing it with the help of the orbital frequency $\omega_0$ of the
circular orbit, or, more conveniently, of the frequency-related
parameter\footnote{From the inverse of
Eq. (\ref{42}) we obtain $r_0$ as a function of $x_0$. For
completeness we give the relations linking both $r_0$ and $R_0$ to the
$x_0$-parameter~:
\begin{eqnarray*}
\frac{m}{r_0\, c^2} &=& x_0 \biggl\{1+\left(1-\frac{\nu}{3}\right)x_0
+ \left(1-\frac{65}{12} \nu \right) x_0^2 \\ &&\quad\quad~
+\left(1+\left[-\frac{10151}{2520}-\frac{41}{192}\pi^2
-\frac{22}{3}\ln\Big(\frac{r_0}{r'_0}\Big)-\frac{44}{9}\lambda\right]\nu
+\frac{229}{36}\nu^2 +\frac{\nu^3}{81}\right)x_0^3 + {\cal
O}\left(x_0^4\,\right)\biggr\}\;,\\
\frac{m}{R_0\, c^2} &=& x_0 \biggl\{1+\left(1-\frac{\nu}{3}\right)x_0
+ \left(\frac{5}{4}-\frac{43}{24} \nu \right) x_0^2 \\ &&\quad\quad~
+\left(\frac{7}{4}+\left[\frac{23759}{5040}-\frac{167}{192}\pi^2
-\frac{44}{9}\lambda\right]\nu +\frac{85}{36}\nu^2
+\frac{\nu^3}{81}\right)x_0^3 + {\cal
O}\left(x_0^4\,\right)\biggr\}\;.
\end{eqnarray*}}

\begin{equation}\label{55}
x_0 \equiv \left(\frac{m\,\omega_0}{c^3}\right)^{2/3}\;.
\end{equation} 
This allows us to write the criterion for stability as $C_0 > 0$,
where $C_0=\frac{m^2}{c^6\,x_0^3}\hat{C}_0$ admits the gauge-invariant
form (the same in all coordinate systems)

\begin{equation}\label{57}
C_0 = 1-6\,x_0 + 14\,\nu\,x_0^2 +
\left(\left[\frac{5954}{35}-\frac{123}{16}\pi^2
-44\lambda\right]\nu-14\nu^2\right)\,x_0^3 + {\cal
O}\left(x_0^4\,\right)\;.
\end{equation} 
This form is more interesting than the coordinate-dependent
expressions (\ref{43}) or (\ref{53}), not only because of its
invariant form, but also because as we see the 1PN term yields exactly
the Schwarzschild result that the innermost stable circular orbit or
ISCO of a test particle (i.e. in the limit $\nu\to 0$) is located at
$x_{\rm ISCO}=1/6$. Thus we find that, at the 1PN order, but for {\it
any} mass ratio $\nu$,

\begin{equation}\label{58}
x_{\rm ISCO}^{\rm 1PN} = \frac{1}{6}\;.
\end{equation}
One could have expected that some deviations of the order of $\nu$
already occur at the 1PN order, but it turns out that only from the
2PN order does one find the occurence of some non-Schwarzschildian
corrections proportional to $\nu$.  At the 2PN order we obtain

\begin{equation}
x_{\rm ISCO}^{\rm 2PN} =
\frac{3}{14\nu}\Bigg(1-\sqrt{1-\frac{14\nu}{9}}~\Bigg)\;.
\end{equation}
For equal masses this gives $x_{\rm ISCO}^{\rm 2PN}\simeq
0.187$. Notice also that the effect of the finite mass corrections is
to increase the frequency of the ISCO with respect to the
Schwarzschild result (i.e. to make it more inward)~: $x_{\rm
ISCO}^{\rm 2PN}=\frac{1}{6}\left[1+\frac{7}{18}\nu+{\cal
O}(\nu^2)\right]$. Finally, at the 3PN order, for equal masses
$\nu=\frac{1}{4}$ and for the value of the ambiguity parameter
$\lambda=-\frac{1987}{3080}$ (equivalent to $\omega_{\rm s}=0$), we
find that according to our criterion all the circular orbits are
stable. More generally, we find that at the 3PN order all orbits are
stable when the mass ratio is $\nu > \nu_c$ where $\nu_c \simeq
0.183$.

Note that the above stability criterion $C_0$ gives an innermost
stable circular orbit, when it exists, that is not necessarily the
same as --- and actually differs from --- the innermost circular orbit
or ICO, which is defined by the point at which the center-of-mass
binding energy of the binary for circular orbits reaches its minimum
\cite{B02ico}. In this respect the present formalism, which is based
on systematic post-Newtonian expansions (without using post-Newtonian
resummation techniques like Pad\'e approximants \cite{DIS98}), differs
from some ``Schwarzschild-like'' methods such as the
effective-one-body approach \cite{BD99} in which the ICO happens to be
also an innermost stable circular orbit or ISCO.

As a final comment, let us note that the use of a {\it truncated}
post-Newtonian series such as Eq. (\ref{57}) to determine the ISCO is
{\it a priori} meaningful only if we are able to bound the neglected
error terms. Furthermore, since we are dealing with a stability
criterion, it is not completely clear that the higher-order
post-Newtonian correction terms, even if they are numerically small,
will not change qualitatively the response of the orbit to the
dynamical perturbation. This is indeed a problem, which cannot be
answered rigorously with the present formalism. However, in the regime
of the ISCO (when it exists), we have seen that $x_0$ is rather small:
$x_0\simeq 0.2$ (this is also approximately the value for the ICO
computed in Ref. \cite{B02ico}), which indicates that the neglected
terms in the truncated series (\ref{57}) should not contribute very
much, because they involve at least a factor $x_0^4\simeq 0.002$. On
the other hand, we pointed out that in the limit $\nu\to 0$ the
criterion $C_0$ gives back the correct {\it exact} result, $x_{\rm
ISCO}^{\nu\to 0} = \frac{1}{6}$. This contrasts with the
gauge-dependent power series (\ref{43}) or (\ref{53}) which give only
some approximate results. Based on these observations, we feel that it
is reasonable to expect that the gauge-invariant stability criterion
defined by Eq. (\ref{57}) is physically meaningful.

\acknowledgments One of us (L.B.) would like to thank Clifford Will
and Thierry Mora for discussions. This work was supported in part by
the National Science Foundation under grant number PHY 00-96522.


\begin{thebibliography}{}
\bibitem{W94} C.M. Will, in the Proc. of the 8th Nishinomiya-Yukawa
Symposium on Relativistic Cosmology, M. Sasaki (ed.), Universal
Acad. Press (1984).
\bibitem{Blivrev} L. Blanchet, Living Rev. Relativity, {\bf 5}, 3
(2002).
\bibitem{DD81a} T. Damour and N. Deruelle, Phys. Lett. {\bf 87}A, 81
(1981).
\bibitem{DD81b} T. Damour and N. Deruelle, C.R. Acad. Sci. Paris {\bf 293},
s\'erie II, 537 (1981).
\bibitem{D82} T. Damour, C.R. Acad. Sci. Paris {\bf 294}, s\'erie II,
1355 (1982).
\bibitem{DS85} T. Damour and G. Sch\"afer, Gen. Relativ. Gravit. {\bf 17}, 879
(1985).
\bibitem{Kop85}S.M. Kopejkin, Astron. Zh. {\bf 62}, 889 (1985).
\bibitem{BFP98}L. Blanchet, G. Faye and B. Ponsot, Phys. Rev. D{\bf 58},
124002 (1998).
\bibitem{IFA01} Y. Itoh, T. Futamase, H. Asada,
Phys.Rev. D {\bf 63},  064038 (2001).
\bibitem{PW02}M.E. Pati and C.M. Will, Phys. Rev. D{\bf 65}, 104008 (2002).
\bibitem{JaraS98} P. Jaranowski and G. Sch\"afer, Phys. Rev. D{\bf 57}, 7274
(1998).
\bibitem{JaraS99} P. Jaranowski and G. Sch\"afer, Phys. Rev. D{\bf 60}, 124003
(1999).
\bibitem{DJS00} T. Damour, P. Jaranowski and G. Sch\"afer,
Phys. Rev. D{\bf 62}, 021501(R) (2000).
\bibitem{DJS01} T. Damour, P. Jaranowski and G. Sch\"afer,
Phys. Rev. D{\bf 63}, 044021 (2001).
\bibitem{DJSdim} T. Damour, P. Jaranowski and G. Sch\"afer,
Phys. Lett. B{\bf 513}, 147 (2001).
\bibitem{BF00} L. Blanchet and G. Faye, Phys. Lett. A{\bf 271}, 58 (2000).
\bibitem{BFeom} L. Blanchet and G. Faye, Phys. Rev. D{\bf 63}, 062005
(2001).
\bibitem{BFreg} L. Blanchet and G. Faye, J. Math. Phys. {\bf 41}, 7675 (2000).
\bibitem{BFregM} L. Blanchet and G. Faye, J. Math. Phys. {\bf 42}, 4391 (2001).
\bibitem{ABF01}V. C. de Andrade, L. Blanchet and G. Faye, 
Class. Quantum Grav. {\bf 18}, 753 (2001).
\bibitem{IW93}B.R. Iyer and C.M. Will, Phys. Rev. Lett. {\bf 70}, 113
(1993).
\bibitem{IW95}B.R. Iyer and C.M. Will, Phys. Rev. D{\bf 52}, 6882 (1995).
\bibitem{GII97}A. Gopakumar, B.R. Iyer and S. Iyer, Phys. Rev. D{\bf
55}, 6030 (1997); Errata: D{\bf 57}, 6562 (1998).
\bibitem{BD88}L. Blanchet and T. Damour, Phys. Rev. D{\bf 37}, 1411
(1988).
\bibitem{B98tail}L. Blanchet, Class. Quantum Grav. {\bf 15}, 113 (1998).
\bibitem{MW02}T. Mora and C.M. Will, submitted to Phys. Rev. D (gr-qc
0208089).
\bibitem{LW90}C.W. Lincoln and C.M. Will, Phys. Rev. D{\bf 42}, 1123
(1990).
\bibitem{BImult} L. Blanchet and B.R. Iyer, work in preparation.
\bibitem{MS79} J. Martin and J.L. Sanz, J. Math. Phys. {\bf 20}, 26 (1979).
\bibitem{B96}L. Blanchet, Phys. Rev. D{\bf 54}, 1417 (1996).
\bibitem{W95}N. Wex, Class. Quantum Grav. {\bf 12}, 983 (1995).
\bibitem{KWW93}L.E. Kidder, C.M. Will and A.G. Wiseman,
Phys. Rev. D{\bf 47}, 3281 (1993).
\bibitem{B02ico}L. Blanchet, Phys. Rev. D{\bf 65}, 124009 (2002);
L. Blanchet, to appear in the Proc. of the 25th Johns Hopkins
Workshop, eds. I. Ciufolini and L. Lusanna (gr-qc 0209089).
\bibitem{DIS98} T. Damour, B.R. Iyer and B.S. Sathyaprakash,
Phys. Rev. D{\bf 57}, 885 (1998).
\bibitem{BD99} A. Buonanno and T. Damour, Phys. Rev. D{\bf 59}, 084006
(1999).
\end{thebibliography}
\end{document}